\RequirePackage[T1]{fontenc}
\documentclass[12pt]{article}

\usepackage[height=8.85in,width=6.45in]{geometry}

\usepackage[utf8]{inputenc}
\usepackage{amsmath}
\usepackage{amssymb}
\usepackage{mathtools}
\numberwithin{equation}{section}
\usepackage{slashed}
\usepackage{braket}
\usepackage[svgnames]{xcolor}
\usepackage[colorlinks,citecolor=DarkGreen,linkcolor=FireBrick]{hyperref}
\usepackage{cite}
\usepackage{graphicx}
\usepackage{tikz}
\usepackage{tikz-cd}
\usepackage{times}
\usepackage{courier}
\usepackage{bm}
\usepackage{subfig}

\usepackage{xcolor}
\usepackage{mdframed}

\renewenvironment{figure}[1][]{
  \begin{originalfigure}[#1]
    \begin{mdframed}[linecolor=black!0,backgroundcolor=black!1]
}{
    \end{mdframed}
  \end{originalfigure}
}

\usepackage{ulem}

\usepackage{colortbl}
\definecolor{dgreen}{rgb}{0, 0.55, 0}
\definecolor{llightyellow}{rgb}{1.0, 0.95, 0.7}
\definecolor{llightblue}{rgb}{0.7, 0.9, 1.0}
\definecolor{llightpink}{rgb}{1.0, 0.85, 0.95}
\definecolor{llightgreen}{rgb}{0.7, 1.0, 0.4}
\colorlet{lightyellow}{llightyellow!50!white}
\colorlet{lightblue}{llightblue!50!white}
\colorlet{lightgreen}{llightgreen!50!white}
\colorlet{lightpink}{llightpink!50!white}

\def\CN{{\mathcal N}}
\def\CO{{\mathcal O}}

\def\CX{{\mathcal X}}

\newcommand{\Z}{\mathbb{Z}}

\newcommand{\T}{\mathcal{T}}

\usepackage{datetime}
\usepackage{dsfont}

\usetikzlibrary{shapes.multipart}
\usetikzlibrary{decorations.pathmorphing}
\tikzset{snake it/.style={decorate, decoration=snake}}

\newcommand{\mS}{\mathsf{S}}
\newcommand{\mT}{\mathsf{T}}
\newcommand{\mU}{\mathsf{U}}
\newcommand{\mV}{\mathsf{V}}

\newcommand{\ssi}{ s^{\sigma}}
\newcommand{\st}{ s^{\tau}}
\newcommand{\hssi}{\widehat{s}^{\sigma}}
\newcommand{\hst}{\widehat{s}^{\tau}}

\newcommand{\KT}{\CN_{\text{KT}}}
\newcommand{\sprimesi}{ s'^{\sigma}}
\newcommand{\sprimet}{ s'^{\tau}}
\newcommand{\sprimeprimesi}{ s''^{\sigma}}
\newcommand{\sprimeprimet}{ s''^{\tau}}
\usetikzlibrary{tikzmark}

\begin{document}

\begin{titlepage}

\begin{flushright}
~
\end{flushright}

\vskip 3cm

\begin{center}

{\Large \bfseries 
Non-Invertible Duality Transformation Between \\\bigskip SPT and SSB Phases
}

\vskip 1cm
Linhao Li$^1$, Masaki Oshikawa$^{1,2}$, and Yunqin Zheng$^{1,2}$
\vskip 1cm

\begin{tabular}{ll}
 $^1$&Institute for Solid State Physics, \\
 &University of Tokyo,  Kashiwa, Chiba 277-8581, Japan\\
 $^2$&Kavli Institute for the Physics and Mathematics of the Universe, \\
 & University of Tokyo,  Kashiwa, Chiba 277-8583, Japan\\
\end{tabular}

\vskip 1cm

\end{center}

\noindent
In 1992, Kennedy and Tasaki constructed a non-local \emph{unitary} transformation that maps between a $\Z_2\times \Z_2$ spontaneously symmetry breaking phase
and the Haldane gap phase, which is a prototypical Symmetry-Protected Topological phase in modern framework,
on an open spin chain.
In this work, we propose a way to define it on a closed chain, by sacrificing unitarity. The operator realizing such a \emph{non-unitary} transformation satisfies \emph{non-invertible} fusion rule, and implements a generalized gauging of the $\Z_2\times \Z_2$ global symmetry. These findings connect the Kennedy-Tasaki transformation to numerous other concepts developed for SPT phases, and opens a way to construct SPT phases systematically using the duality mapping.

\end{titlepage}

\setcounter{tocdepth}{3}
\tableofcontents

\section{Introduction}

Symmetry-Protected Topological (SPT) phases are one of the central issues in contemporary quantum many-body physics.
While the concept of SPT phases was established~\cite{2009PhRvB..80o5131G, PhysRevB.81.064439, PhysRevB.85.075125, Chen:2011pg} after the discovery of topological insulators~\cite{PhysRevLett.95.226801,PhysRevLett.95.146802}---the free electron version of SPT phases, a prototypical example of the bosonic SPT phases \cite{PhysRevLett.50.1153}---the Haldane gap phase, was found much earlier.
Although Haldane's initial prediction was only that the Heisenberg antiferromagnetic chains with integer spins
are ``massive'' (with a non-zero excitation gap and exponentially decaying correlation functions),
it was gradually recognized that the Haldane gap phase has various exotic properties.

In \cite{PhysRevLett.59.799}, Affleck, Kennedy, Lieb, and Tasaki constructed exact ground state wavefunctions for a certain generalized Heisenberg antiferromagnetic
chains, which are now called AKLT states/models.
We can naturally see from its construction, the AKLT model on an open chain exhibits free fractionalized edge spins~\cite{kennedy1990exact}.
Moreover, although there is no long-range order in the conventional sense that can be detected by correlation functions of
local operators, the AKLT state can be characterized by a non-local \textit{string order parameter} \cite{PhysRevB.40.4709}.
It was also confirmed numerically that these properties are not specific to the AKLT model but are characteristics of the
\textit{Haldane gap phase} including the ground state of the standard Heisenberg antiferromagnetic chain with $S=1$.

Kennedy and Tasaki~\cite{1992CMaPh.147..431K, PhysRevB.45.304}
demonstrated that these two apparently unrelated features of the $S=1$ Haldane gap phase can be understood as consequences of
the spontaneous breaking of a \textit{hidden $\Z_2\times \Z_2$ symmetry}.
That is, they showed that many $S=1$ spin Hamiltonians with short-range interactions on an open chain
are mapped to Hamiltonians with short-range interactions
by a non-local unitary transformation now called Kennedy-Tasaki (KT) transformation.
While the Kennedy-Tasaki transformation was introduced specifically for $S=1$ chains in a complicated way in the original literature,
a simple compact expression, which is valid for any integer spin, was found as
\begin{eqnarray}
\label{eq:KToriginal}
U_{\text{KT}}=\prod_{i>j} \exp\left(i\pi S^{z}_i S^x_j\right)
\end{eqnarray}
by one of the authors of the present work~\cite{oshikawa1992hidden}.
It maps the spin operators as follows:
\begin{eqnarray}
U_{\text{KT}} S^x_j {U_{\text{KT}}}^\dagger &=& S^x_j e^{i \pi \sum_{k<j} S^x_k },
\label{eq:Sx_KT}
\\
U_{\text{KT}} S^z_j {U_{\text{KT}}}^\dagger &=& e^{i\pi \sum_{k>j} S^z_k} S^z_j ,
\label{eq:Sz_KT}
\\
U_{\text{KT}} S^y_j {U_{\text{KT}}}^\dagger &=& e^{i\pi \sum_{k>j} S^z_k} S^y_j e^{i \pi \sum_{k<j} S^x_k } .
\label{eq:Sy_KT}
\end{eqnarray}
While the local spin operators are mapped to non-local operators, many of the spin chain Hamiltonians of
interest are sums of local (i.e. defined over a short range) quadratic forms of spin operators
and thus are mapped to another Hamiltonian with short-range interactions by the Kennedy-Tasaki transformation.
The Hamiltonian obtained in this way generally possesses a dihedral symmetry of global spin rotations
($\pi$-rotation about $x,y,$ and $z$ axes),
which is isomorphic to the $\Z_2\times \Z_2$.
A spontaneous breaking of this ``hidden'' $\Z_2\times \Z_2$ symmetry
implies the long-range string order in the original system, as well as
a four-fold ground-state degeneracy of the open chain signaling the existence of the 
fractional $S=1/2$ edge spins \cite{TKennedy_1990}.

It is remarkable that, the recognition of the ``Haldane gap phase'' as a nontrivial phase,
which cannot be characterized by any local order parameter
but is distinct from a trivial phase, was established by early 1990s even though
the clear concept of the SPT phases was missing.
In fact, in the early days, the Haldane gap phase was sometimes called ``topological'' or ``topological ordered''
without a clear definition~\cite{SMGirvin_1989}.
To be specific, what was lacking at that time was the recognition that a certain symmetry is required for
the Haldane gap phase to be distinct from the trivial phase.
In retrospect, however, the global $\Z_2 \times \Z_2$ symmetry could have been
identified as a symmetry protecting the Haldane gap phase, since it is the
condition for the Kennedy-Tasaki transformation to give a Hamiltonian with short-range interactions~\cite{PhysRevB.85.075125}.

A comment is in order on the terminology.
Although the Kennedy-Tasaki transformation~\eqref{eq:KToriginal} is a highly non-local transformation,
the symmetry operators
\begin{equation}
    R^\alpha = e^{i \pi \sum_j S^\alpha_j} \;\;\; (\alpha=x,y,z) ,
    \label{eq:defR}
\end{equation}
implementing the $\Z_2 \times \Z_2$ transformation are invariant under the Kennedy-Tasaki transformation:
\begin{equation}
  U_{\text{KT}} R^\alpha U_{\text{KT}}^\dagger = R^\alpha  ,
  \label{eq:R_UKT}
\end{equation}
because the eigenvalues of $e^{i \pi S^\alpha_j}$ are $\pm 1$ and $e^{i\pi S^\alpha_j} = e^{-i\pi S^\alpha_j}$, for
$\alpha = x, y, z$~\cite{Oshikawa2012}.
In this sense, the $\Z_2 \times \Z_2$ symmetry is not exactly ``hidden'', since it is (a subset of) the symmetry
of the original Hamiltonian with the open boundary condition.
However, it is certainly appropriate that the SPT phase corresponds
to the hidden (spontaneous) breaking of the symmetry, because the symmetry is not spontaneously broken
in the SPT phase before applying the Kennedy-Tasaki transformation.

Since the discovery of the concept of the SPT phases, significant progress has been made on many fronts.
In particular, systematic classifications in general dimensions have been studied, uncovering the deep relation to
topological quantum field theory and algebraic topology \cite{Chen:2011pg, Freed:2016rqq, Kitaev:2009mg, Kapustin:2014dxa, Kapustin:2014gma, Kapustin:2014tfa}.
Given the history, it would be worthwhile to revisit the Kennedy-Tasaki transformation, which played a significant role in understanding
the Haldane gap phase 30 years ago, from the modern perspective.
It is indeed the goal of the present paper.

It must be mentioned that various properties, generalizations and applications of the vanilla Kennedy-Tasaki transformation have been explored. The hidden symmetry breaking order in models of spin-$1/2$ and higher integer spin was studied in \cite{PhysRevB.46.3486, takada1992nonlocal, 1995EL.....30..493S, 1996PhRvB..54.4038S}. The relation between the hidden symmetry breaking order and SPT order with a broad class of symmetry, such as $\Z_N\times \Z_N$, was discussed in \cite{PhysRevB.78.094404,2013PhRvB..88h5114E,PhysRevB.88.125115}. Moreover, the Kennedy-Tasaki transformation can disentangle the twofold degeneracy entanglement spectrum of spin-1 Heisenberg chain \cite{PhysRevB.83.104411,PhysRevB.89.125112}. Besides, the phase diagram of a $(1+1)$d model, which is defined by interpolating between the spin-1 bilinear-biquadratic chain and its Kennedy-Tasaki dual, was discussed in \cite{yang2022duality}.

However, most of the discussions of the (generalized) Kennedy-Tasaki transformation so far focused on open boundary conditions;
to the best of the authors' knowledge, the Kennedy-Tasaki transformation on closed chains has not been explored.
In fact, there are several apparent difficulties in defining on closed chains. 
\begin{enumerate}
    \item The highly non-local unitary operator \eqref{eq:KToriginal} depends on the ordering of the sites. On an open chain, the ordering is well-defined. However, for the closed chain, there is no consistent ordering. The site $i$ to the left of another site $j$ can also be viewed as to the right of $j$ by going around the ring. 
    \item An SPT phase on a closed chain only has a single non-degenerate ground state. However, an SSB phase has multiple degenerate ground states. Hence the two Hamiltonians on a closed chain can not be mapped to each other via a unitary transformation. 
\end{enumerate}

Nevertheless, there are several reasons to consider this transformation beyond on open chains.
First, it is theoretically demanding that a physically well-defined operation should be applicable to all boundary conditions.
Second, twisting the boundary condition on a closed chain is useful to characterize the SPT phases.
In order to exploit the twisted boundary conditions as a probe, it is desirable to construct the Kennedy-Tasaki transformation on closed chains.
Third, although Kennedy-Tasaki transformation has a rather compact expression~\eqref{eq:KToriginal}, its physical interpretation is
not quite clear. This is perhaps one of the reasons why this intriguing transformation has not been generalized beyond the Haldane gap phase
($\mathbb{Z}_2 \times \mathbb{Z}_2$-protected SPT phase in one dimension).
As we will find later in this work, by investigating the transformation on a closed chain,
we are able to find a more transparent physical interpretation from a modern viewpoint:
it is simply gauging the $\Z_2\times \Z_2$ symmetry with certain twist.

In this work, we propose that the Kennedy-Tasaki transformation can be defined on closed chains by sacrificing unitarity or by expanding the Hilbert space by
including ``twist'' sectors corresponding to different boundary conditions. 
We make two attempts to define the Kennedy-Tasaki transformation on a ring. 
\begin{enumerate}
    \item The first attempt is to define the Kennedy-Tasaki transformation for spin-1 system by naively implementing \eqref{eq:Sx_KT}, \eqref{eq:Sz_KT} and \eqref{eq:Sy_KT} on a ring. 
    \item The second attempt is to propose a non-unitary transformation $\KT$ acting on a ring where each unit cell contains two spin-$\frac{1}{2}$'s.
     For convenience, we also call $\KT$ the Kennedy-Tasaki transformation.
    Similar generalizations of the Kennedy-Tasaki transformation to the spin-$\frac{1}{2}$ systems with open boundary condition were constructed
    earlier~\cite{PhysRevB.46.3486,takada1992nonlocal}. Our construction is also inspired by the recent work~\cite{PhysRevB.96.125104,Moradi:2022lqp,Moradi:2023dan}. 
     We will show that $\KT$ satisfies the desired properties: it maps an $\Z_2\times \Z_2$ SSB phase to an $\Z_2\times \Z_2$ SPT phase both on a closed and open chain.
\end{enumerate}
The Kennedy-Tasaki transformation from the two attempts will be shown to be equivalent on a ring. In particular, the transformations for both spin-1 and two spin-$\frac{1}{2}$ systems are non-unitary transformations and satisfy the non-invertible fusion rules, which is a generalization of the famous Kramers-Wannier duality transformation. Such duality  transformations have been extensively discussed in recent years, both in $(1+1)$d\cite{cobanera2011bond,Chang:2018iay,Bhardwaj:2017xup,Aasen:2016dop, Aasen:2020jwb,Thorngren:2019iar,Thorngren:2021yso, PhysRevB.104.125418,Lootens:2022avn} and in higher dimensions\cite{Kaidi:2021xfk,Kaidi:2022cpf,Kaidi:2022uux,Choi:2021kmx,Choi:2022jqy,Choi:2022rfe,Choi:2022zal, Cordova:2022ieu, Bhardwaj:2022yxj, Bhardwaj:2022lsg, Bartsch:2022mpm, Bhardwaj:2022kot,Bartsch:2022ytj}. When the system is invariant under the duality transformation, the operator $\KT$ becomes a non-invertible symmetry of the system.

Although the two attempts can be shown to be equivalent, the construction in the second attempt is more convenient to manipulate because the degrees of freedom charged under two $\Z_2$'s are decoupled. The first and the second spin-$\frac{1}{2}$ are charged under separate $\Z_2$'s respectively. Moreover, the decoupling between degrees of freedom admits a more convenient interpretation of twisted gauging, similar to the Kramers-Wannier duality transformation which implements gauging of the $\Z_2$ global symmetry.  
This interpretation also facilitates the construction of new models with interesting topological features. In an upcoming work \cite{Li:2023knf}, we will apply the Kennedy-Tasaki transformation to systematically construct a series of gapless SPT phases that have been recently explored in \cite{scaffidi2017gapless, 2019arXiv190506969V,Thorngren:2020wet,Li:2022jbf, Wen:2022lxh} and uncover new ones.

This paper is organized as follows. In Section \ref{sec:Kramers-Wannier}, we systematically review the Kramers-Wannier (KW) transformation on closed and open chains respectively, as a preparation for a more complicated Kennedy-Tasaki transformation. The Kramers-Wannier transformation on a closed chain is well-known to be non-unitary and the operators implementing the Kramers-Wannier transformation satisfy the non-invertible fusion rule.
However, on an open chain with the free boundary condition, the Kramers-Wannier transformation can be defined as a unitary transformation, as it was pointed out in Ref.~\cite{Nussinov:2012nz}.
In Section \ref{sec:KTspin1ring}, we define the Kennedy-Tasaki transformation for spin-1 systems on a ring, and find that it is non-unitary and obeys the non-invertible fusion rule. In Section \ref{sec:fieldtheory}, we motivate that the Kennedy-Tasaki transformation implements a twisted gauging, via field theory formulation. In Section \ref{sec:KTclosedbdy}, we define the Kennedy-Tasaki transformation for spin-$\frac{1}{2}$ systems on a ring, and explore nice properties in parallel with the Kramers-Wannier transformations. These properties coincide with those in Section \ref{sec:KTspin1ring}. In Section \ref{sec:KTopenbdy}, we place the Kennedy-Tasaki transformation for spin-$\frac{1}{2}$ systems on an interval, and find that it becomes a unitary operator. In Section \ref{sec:gappedSPT}, we explain how to construct the typical representative model of SPT---the cluster model, using the Kennedy-Tasaki transformation. Finally in Section \ref{sec:projtospin1}, we prove the equivalence between the Kennedy-Tasaki transformations for spin-1 and spin-$\frac{1}{2}$ systems.

\section{Kramers-Wannier transformation}
\label{sec:Kramers-Wannier}

To prepare for the reformulation of the Kennedy-Tasaki transformation, we first discuss the Kramers-Wannier transformation.
While the Kramers-Wannier transformation has been known for many years, here we shall formulate it precisely,
with an emphasis on modern concepts such as non-invertible fusion rules and mapping between symmetry and twist sectors.
This is not only because it is a precursor of the Kennedy-Tasaki transformation as a nonlocal duality mapping;
we will reformulate the Kennedy-Tasaki transformation \emph{based on} the Kramers-Wannier transformation in later sections.


The Kramers-Wannier transformation was initially conceived as a duality mapping between a higher temperature
and a lower temperature of the two-dimensional classical Ising model~\cite{Kramers-Wannier}.
The simple assumption of the existence of the single phase transition between the disordered
and ordered phases, combined with the Kramers-Wannier transformation, determines the critical temperature
on the square lattice uniquely. 
As a typical example of the general correspondence between classical statistical systems in 2 dimensions
and quantum many-body systems in 1 spatial dimension,
the quantum transverse-field Ising chain defined by the Hamiltonian
\begin{eqnarray}
H_{\text{Ising}}^h = - \sum_i \left( \sigma^z_{i-1}\sigma^z_{i} + h \sigma^x_{i}\right).
\label{eq:IsingH}
\end{eqnarray}
is a counterpart of the two-dimensional classical Ising model.
The Kramers-Wannier transformation can be also defined for the quantum spin model~\eqref{eq:IsingH} in one spatial dimension~\cite{Kogut-RMP1979}.
(In fact, the Kramers-Wannier transformation is applicable to more general systems and not limited to the particular model, as we will see later.)
However, there are subtleties related to the boundary conditions of the system, as we will discuss below.

Recently, the Kramers-Wannier transformation has been also reformulated from the modern viewpoint.
The transverse-field Ising chain~\eqref{eq:IsingH} has a global $\Z_2$ symmetry, which is generated by the simultaneous flip of spin at every site.
This is a typical example of an ``on-site symmetry'' because the symmetry generator is a product of single-site operators.
Naturally, such a symmetry can be gauged by introducing local gauge transformation (local spin flips).
The Kramers-Wannier transformation may be identified with such a ``gauging'' operation of the $\Z_2$ symmetry\cite{Thorngren:2021yso,Thorngren:2019iar,Aasen:2016dop,Aasen:2020jwb,Bhardwaj:2017xup,Chang:2018iay}
.

In terms of field theory, the gauging is understood as a topological manipulation corresponding to an insertion of a line defect $\CN$ in $(1+1)$-dimensional space-time,
to obtain the new system $\T/\Z_2$ from a given system $\T$ with the global $\Z_2$ symmetry.
The defect satisfies the fusion rule of ``Ising-category''
\begin{eqnarray}\label{eq:IsingFusionrule}
\CN^{\dagger}\times \CN = 1+ U, \hspace{1cm} \widehat{U}\times \CN= \CN\times U = \CN, \hspace{1cm} U\times U=1.
\end{eqnarray}
where $U$ and $\widehat{U}$ are the topological line defects that generate the $\Z_2$ symmetry in $\T$ and the $\Z_2$ symmetry in $\T/\Z_2$. 
This fusion rule implies that $\CN$ lacks its inverse; the Kramers-Wannier transformation is thus ``non-invertible''.
While this statement may look rather abstract, in the following we will define the Kramers-Wannier transformation carefully on the lattice,
with a particular emphasis on subtleties concerning the symmetry and twist sectors.
Our discussion leads to a more precise version of the fusion rule, and elucidates its physical meaning.
The prior knowledge of the fusion rule~\eqref{eq:IsingFusionrule} is not necessary to follow the discussion in this Section.


\subsection{Kramers-Wannier transformation on a closed chain}
\label{sec:KWtransformation_closed}

Let us consider a spin chain with $L$ sites. Each site supports one spin-$\frac{1}{2}$ spanning a two dimensional local Hilbert space $\ket{s_i}$, where $s_i=0,1$ and $i=1, ..., L$. The state can be acted upon by spin measurement and spin flip Pauli operators in the standard way, 
\begin{eqnarray}\label{eq:Paulioper}
\sigma^z_{i}\ket{s_i}= (-1)^{s_i} \ket{s_i}, \hspace{1cm} \sigma^x_{i} \ket{s_i} = \ket{1-s_i}.
\end{eqnarray}
We also assume that the spin system has an on-site $\Z_2$ global symmetry, generated by
\begin{eqnarray}\label{eq:Z2generatoronsite}
U= \prod_{i=1}^L \sigma^x_i
\end{eqnarray}
which flips the spins on every site simultaneously. The states can be organized into eigenstates of $U$ as $\Z_2$ even state with the eigenvalue of $U$ to be $(-1)^u = 1$, i.e. $u=0$, and $\Z_2$ odd state with the eigenvalue of $U$ to be $(-1)^u=-1$, i.e. $u=1$. Moreover, one can also use the $\Z_2$ symmetry to twist the boundary condition of the spins as
\begin{equation}
\ket{s_{i+L}}= (\sigma^x_i)^t\ket{s_{i}}= \ket{s_{i}+t},
\label{eq:Ising_bc}
\end{equation}
hence the spins obey either periodic boundary condition (PBC), i.e. $t=0$ or twisted boundary condition (TBC), i.e. $t=1$. In summary, one can organize the Hilbert space into four symmetry-twist sectors, labeled by $(u,t)\in \{0,1\}^2$.

It is also useful to define a set of ``dual" spin-$\frac{1}{2}$'s on the links between sites. We use half-integers to label the position of links, and dual spins are labeled by $\widehat{s}_{i-\frac{1}{2}}$'s, where $i=1, ..., L$. They also span local Hilbert spaces on the links $\ket{\widehat{s}_{i-\frac{1}{2}}}$. The states are acted upon by Pauli operators $\tau^z_{i-\frac{1}{2}}$ and  $\tau^x_{i-\frac{1}{2}}$, similar as \eqref{eq:Paulioper}, 
\begin{eqnarray}
\tau^z_{i-\frac{1}{2}} \ket{\widehat{s}_{i-\frac{1}{2}}} = (-1)^{\widehat{s}_{i-\frac{1}{2}}}\ket{\widehat{s}_{i-\frac{1}{2}}}, \hspace{1cm} \tau^x_{i-\frac{1}{2}} \ket{\widehat{s}_{i-\frac{1}{2}}} = \ket{1-\widehat{s}_{i-\frac{1}{2}}}.
\end{eqnarray}
The dual on-site $\Z_2$ global symmetry acting on the links, generated by 
\begin{eqnarray}
\widehat{U} = \prod_{i=1}^{L} \tau_{i-\frac{1}{2}}^x.
\end{eqnarray}
Likewise, the dual Hilbert space can also be organized into four sectors labeled by
$(\widehat{u}, \widehat{t}) \in \{0,1\}^2$. 
Note that the spins and the dual spins do not exist as independent degrees of freedom simultaneously, rather, one determines the other by the Kramers-Wannier transformation.

Following \cite{Aasen:2016dop}, the Kramers-Wannier transformation would be defined
as the operator $\CN$ acting on the Hilbert space, in terms of the matrix elements
\begin{eqnarray}
\bra{\{\widehat{s}_{i-\frac{1}{2}} \}}
\CN \ket{\{s_i\}} &\sim& \frac{1}{2^{L/2}} (-1)^{\sum_j s_{j} ( \widehat{s}_{j - \frac{1}{2}} + \widehat{s}_{j+\frac{1}{2}} ) } 
\label{eq:KWinf1}
\\
&\sim& \frac{1}{2^{L/2}} (-1)^{\sum_j \left( s_{j-1} + s_j \right)  \widehat{s}_{j - \frac{1}{2}}  } ,
\label{eq:KWinf2}
\end{eqnarray}
on an infinite chain, where the above two expressions are equivalent.
However, on the finite ring, the summation should be limited to $L$ sites (or dual sites), and
the boundary conditions should be carefully examined.

Let us start from the expression~\eqref{eq:KWinf1} and limit the summation to $\sum_{j=1}^L$.
The last term in the sum contains $s_L \widehat{s}_{L+1/2}$.
If we are to define the dual spins $\widehat{s}$ on the half-integer sites $1/2,3/2,\ldots,L-1/2$, $\widehat{s}_{L+\frac{1}{2}}$
should be replaced by $\widehat{s}_{1/2}+\hat{t}$ (modulo 2), using the boundary condition $\hat{t}=0,1$ for the dual spin.
Then the Kramers-Wannier transformation on the ring seems to be given by
\begin{eqnarray}
\bra{\{\widehat{s}_{i-\frac{1}{2}} \}}
\CN \ket{\{s_i\}} \sim \frac{1}{2^{L/2}} (-1)^{\sum_{j=1}^L s_{j} \widehat{s}_{j-\frac{1}{2}} + \sum_{j=1}^{L-1} s_j \widehat{s}_{j+\frac{1}{2}} 
+ s_L \widehat{s}_{\frac{1}{2}} +  \widehat{t}s_L} .
\label{eq:KWguess1}
\end{eqnarray}
On the other hand, starting from Eq.~\eqref{eq:KWinf2} and limiting the summation to $\sum_{j=1}^L$,
we find the ``boundary term'' $s_0 \widehat{s}_{\frac{1}{2}}$.
Replacing $s_0$ with $s_L + t$, we find 
\begin{eqnarray}
\bra{\{\widehat{s}_{i-\frac{1}{2}} \}}
\CN \ket{\{s_i\}} \sim \frac{1}{2^{L/2}} (-1)^{\sum_{j=1}^L s_{j} \widehat{s}_{j-\frac{1}{2}} + \sum_{j=1}^{L-1} s_j \widehat{s}_{j+\frac{1}{2}} 
+ s_L \widehat{s}_{\frac{1}{2}} +  t \widehat{s}_{\frac{1}{2}}} .
\label{eq:KWguess2}
\end{eqnarray}
In this way, we can ``derive'' two different (and inequivalent) expressions for the Kramers-Wannier transformation on the ring.

It turns out that the correct expression for the Kramers-Wannier transformation on the ring includes both boundary factors
appearing in Eqs.~\eqref{eq:KWguess1} and~\eqref{eq:KWguess2}, and is given as 
\begin{eqnarray}
\label{eq:defKW}
\bra{\{\widehat{s}_{i-\frac{1}{2}} \}}
\CN \ket{\{s_i\}} = \frac{1}{2^{L/2}} (-1)^{\sum_{j=1}^L s_{j} \widehat{s}_{j-\frac{1}{2}} + \sum_{j=1}^{L-1} s_j \widehat{s}_{j+\frac{1}{2}} 
+ t \widehat{s}_{\frac{1}{2}} +  \widehat{t}s_L} .
\end{eqnarray}
We will confirm that this is the appropriate definition of the Kramers-Wannier transformation by explicit calculations. In particular, the boundary terms can be fixed by matching how the symmetry-twist sectors are mapped from gauging $\Z_2$, as in \cite{Hsieh:2020uwb}. 
The above expression can be also written as either
\begin{eqnarray}
\bra{\{\widehat{s}_{i-\frac{1}{2}} \}}
\CN \ket{\{s_i\}} = \frac{1}{2^{L/2}} (-1)^{\sum_{j=1}^L \left( s_{j-1} + s_j \right)  \widehat{s}_{j - \frac{1}{2}} + \widehat{t}s_L} ,
\label{eq:defKW_hatt}
\end{eqnarray}
where $s_0 = s_L + t$ is understood, or
\begin{eqnarray}
\bra{\{\widehat{s}_{i-\frac{1}{2}} \}}
\CN \ket{\{s_i\}} = \frac{1}{2^{L/2}} (-1)^{\sum_{j=1}^L s_{j} \left( \widehat{s}_{j - \frac{1}{2}} + \widehat{s}_{j+\frac{1}{2}} \right)+ t \widehat{s}_{\frac{1}{2}}} ,
\label{eq:defKW_t}
\end{eqnarray}
where $\widehat{s}_{L+\frac{1}{2}}=\widehat{s}_{\frac{1}{2}}+\widehat{t}$ is understood.
Eqs.~\eqref{eq:defKW}, \eqref{eq:defKW_hatt}, and \eqref{eq:defKW_t} are equivalent, while
they contain an extra factor compared to the naive versions~\eqref{eq:KWguess1} or \eqref{eq:KWguess2}.

The operator $\CN$ acts on the Hilbert space of the entire system at a certain ``time slice''.
Thus it corresponds to a defect line parallel to the spatial axis
in the $(1+1)$-dimensional space-time.
We remark that by exchanging the role of space and time, the operator $\CN$ can be interpreted as a defect in the Hilbert space. This point of view was more often adopted in the recent discussions of non-invertible defects and their fusion rules \cite{Kaidi:2021xfk, Choi:2021kmx,Choi:2022zal}. We will only work with operator $\CN$ acting on the Hilbert space in this work. Moreover, the $\CN$ operator defined this way is independent of the underlying Hamiltonian. Instead one can use $[H, \CN]\ket{\{s_i\}} =0$ to constrain the possible Hamiltonians which are self-dual under the Kramers-Wannier transformation.

\subsection{Fusion rules}
\label{sec:KWfusion}

We proceed to discuss the fusion rule involving $\CN$ and  $U$.
Since we have defined the duality and symmetry defects as operators $\CN$ and $U$,
the ``fusion'' is simply given as a product of the operators.
We start from a general state of the original spins
\begin{eqnarray}
\ket{\psi} = \sum_{\{s_i\}} \psi_{\{s_i\}} \ket{\{s_i\}}
\end{eqnarray}
where $\psi_{\{s_i\}}$ is the wavefunction of the spin variables. 

Let us first consider $\CN\times U$. 
\begin{equation}
\begin{split}
    \CN\times U \ket{\psi} &=  \CN  \sum_{\{s_i\}} \psi_{\{1-s_i\}} \ket{\{s_i\}}\\
    &= \frac{1}{2^{L/2}}\sum_{\{\widehat{s}_{i-\frac{1}{2}}\},\{s_i\}} \psi_{\{1-s_i\}}   (-1)^{\sum_{j=1}^L (s_{j-1}+ s_{j})\widehat{s}_{j-\frac{1}{2}} + \widehat{t}s_L} \ket{\{\widehat{s}_{i-\frac{1}{2}}\}}\\
    &= \frac{1}{2^{L/2}}\sum_{\{\widehat{s}_{i-\frac{1}{2}}\},\{s_i\}} \psi_{\{s_i\}}   (-1)^{\sum_{j=1}^L (s_{j-1}+ s_{j})\widehat{s}_{j-\frac{1}{2}} + \widehat{t}(1-s_L)} \ket{\{\widehat{s}_{i-\frac{1}{2}}\}}\\
    &= (-1)^{\widehat{t}}\CN \ket{\psi}.
\end{split}
\end{equation}
This implies the fusion rule
\begin{eqnarray}\label{eq:NUfusion}
\CN \times U = (-1)^{\widehat{t}}\CN.
\end{eqnarray}
This fusion rule is slightly different from the standard fusion rule in the Ising fusion category \eqref{eq:IsingFusionrule}, by the additional factor $(-1)^{\widehat{t}}$. Such a factor can be traced back to the additional term  $\widehat{t}s_L$ in \eqref{eq:defKW}. Here, we would like to argue that $(-1)^{\widehat{t}}$ makes sense.

The fusion rule~\eqref{eq:NUfusion} implies that the spin-flip parity $u$ of the original spins is linked to the boundary condition $\widehat{t}$ of the dual spins.
That is, for any parity eigenstate 
\begin{eqnarray}
    U | \Psi \rangle = (-1)^u |\Psi \rangle ,
\end{eqnarray}
it follows from Eq.~\eqref{eq:NUfusion} that
\begin{eqnarray}
   (-1)^{\widehat{t}} \left( \CN |\Psi \rangle \right) = (-1)^u \left( \CN |\Psi \rangle \right) ,
\end{eqnarray}
namely
\begin{eqnarray}
    \widehat{t} = u .
\label{eq:hatt_u}
\end{eqnarray}

\begin{figure}
    \centering
    \begin{tikzpicture}
	\draw[thick] (0,0) circle (1);
	\draw[thick, color=red] (0,0) circle (0.5);
	\draw[thick, fill=black!90] (0,0) circle (1pt);
	\node() at (0.15,-0.15){$\sigma$} ;
	\node()[color=red] at (-0.2,0.65){$U$} ;
	\node() at (1.3,0){$\CN$} ;
	\draw[thick] (2.5,1) -- node[sloped, anchor=center, above] {=}  (2.51,1.005);
	
	\node() at (3.2,1.5){$-$} ;
	\draw[thick] (4.5,1.5) circle (1);
	\draw[thick, fill=black!90] (4.5,1.5) circle (1pt);
	\node() at (4.65,1.35){$\sigma$} ;
	\node() at (5.8,1.5){$\CN$} ;

	\draw (2.2,-1.2) -- node[sloped, anchor=center, above] {=}  (2.21,-1.205);

	\draw[thick] (4,-1.5) circle (1);
	\draw[thick, fill=black!90] (4,-1.5) circle (1pt);
	\node() at (4.15,-1.65){$\sigma$} ;
	\node()[color=red] at (3.7,-1.05){$U$} ;
	\node() at (5.3,-1.5){$\CN$} ;
	\draw[thick, color=red] (4.94 ,-1.25) arc (90:270:0.3);
	\draw[thick, color=red] (3.75 ,-0.56) arc (180:360:0.3);
	\draw[thick, color=red] (3.06 ,-1.75) arc (280:440:0.3);
	\draw[thick, color=red] (4.25 ,-2.44) arc (0:180:0.3);

	\node() at (6,-1.5){=} ;
	\draw[thick] (7.5,-1.5) circle (1);
	\draw[thick, fill=black!90] (7.5,-1.5) circle (1pt);
	\node() at (7.65,-1.65){$\sigma$} ;
	\node() at (8.8,-1.5){$\CN$} ;

	\end{tikzpicture}
    \caption{Shrink the $U$ operator inwards and expanding $U$ outwards yield opposite signs. This means that $\CN\ket{\psi}$ for $\Z_2$ odd $\ket{\psi}$ vanishes in the untwisted sector.  }
    \label{fig:untwisted}
\end{figure}
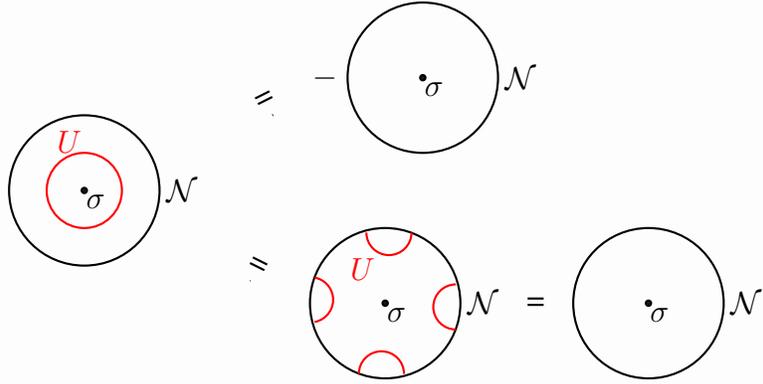

It is useful to see \eqref{eq:NUfusion} in a diagrammatic way. Let us justify this in the Ising CFT. In the Ising CFT, there are three local primary operators, the trivial operator, the energy operator $\varepsilon$ and the spin operator $\sigma$. The spin operator is $\Z_2$ odd, while the energy operator is $\Z_2$ even. Let us first prepare a $\Z_2$ odd state $\ket{\sigma}$ in the untwisted sector, by acting $\sigma$ on the vacuum state $\ket{0}$. In the radial quantization picture, the state is represented by placing $\sigma$ at the origin. Let us act $U$ and $\CN$ on the state, by wrapping $U$ and $\CN$ subsequently around $\sigma$. We then shrink the $U$ operator in two different ways as shown in Figure \ref{fig:untwisted}. Shrinking $U$ inward means acting $U$ on $\ket{\sigma}$, and we obtain a minus sign since $\sigma$ is $\Z_2$ odd. One can also expand the $U$ outwards. By applying the F-move several times\cite{Chang:2018iay,Aasen:2016dop, Aasen:2020jwb,Moore:1988qv, Frohlich:2004ef, Frohlich:2006ch, Frohlich:2009gb}, and one finds that $U$ can be absorbed by $\CN$. Hence the $\CN \ket\psi$ vanishes in the untwisted sector.

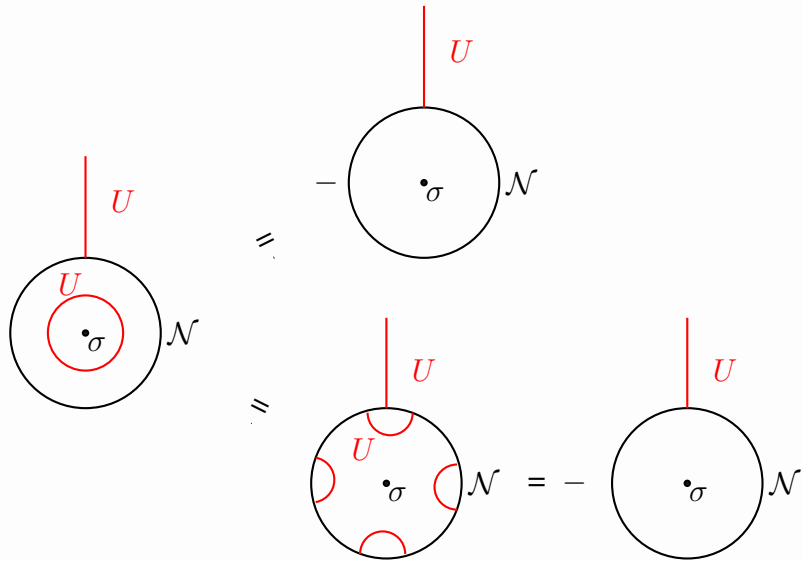
\begin{figure}
    \centering
    \begin{tikzpicture}
	\draw[thick] (0,0) circle (1);
	\draw[thick, color=red] (0,0) circle (0.5);
	\draw[thick, fill=black!90] (0,0) circle (1pt);
	\node() at (0.15,-0.15){$\sigma$} ;
	\node()[color=red] at (-0.2,0.65){$U$} ;
	\node() at (1.3,0){$\CN$} ;
	\draw[thick] (2.5,1) -- node[sloped, anchor=center, above] {=}  (2.51,1.005);
	\node(b01) at (0,0.85){} ;
	\node(b02) at (0,2.5){} ;
	\node[red] at (0.5,1.75){$U$} ;
	\path[thick, color=red] (b01) edge (b02);
	
	\node() at (3.2,2){$-$} ;
	\draw[thick] (4.5,2) circle (1);
	\draw[thick, fill=black!90] (4.5,2) circle (1pt);
	\node() at (4.65,1.85){$\sigma$} ;
	\node() at (5.8,2){$\CN$} ;
	\node(b03) at (4.5,2.85){} ;
	\node(b04) at (4.5,4.5){} ;
	\node[red] at (5,3.75){$U$} ;
	\path[thick, color=red] (b03) edge (b04);
	
	\draw (2.2,-1.2) -- node[sloped, anchor=center, above] {=}  (2.21,-1.205);

	\draw[thick] (4,-2) circle (1);
	\draw[thick, fill=black!90] (4,-2) circle (1pt);
	\node() at (4.15,-2.15){$\sigma$} ;
	\node()[color=red] at (3.7,-1.55){$U$} ;
	\node() at (5.3,-2){$\CN$} ;
	\draw[thick, color=red] (4.94 ,-1.75) arc (90:270:0.3);
	\draw[thick, color=red] (3.75 ,-1.06) arc (180:360:0.3);
	\draw[thick, color=red] (3.06 ,-2.25) arc (280:440:0.3);
	\draw[thick, color=red] (4.25 ,-2.94) arc (0:180:0.3);
	\node(b05) at (4,-1.15){} ;
	\node(b06) at (4,0.35){} ;
	\node[red] at (4.5,-0.5){$U$} ;
	\path[thick, color=red] (b05) edge (b06);

	\node() at (6,-2){=} ;
	\node() at (6.5,-2){$-$} ;
	\draw[thick] (8,-2) circle (1);
	\draw[thick, fill=black!90] (8,-2) circle (1pt);
	\node() at (8.15,-2.15){$\sigma$} ;
	\node() at (9.3,-2){$\CN$} ;
    \node(b07) at (8,-1.15){} ;
	\node(b08) at (8,0.35){} ;
	\node[red] at (8.5,-0.5){$U$} ;
	\path[thick, color=red] (b07) edge (b08);

	\end{tikzpicture}
    \caption{Shrink the $U$ operator inwards and expanding $U$ outwards yield the same minus sign. This is consistent with the fact that $\CN\ket{\psi}$ for $\Z_2$ odd $\ket{\psi}$ is in general non-vanishing in the twisted sector. }
   \label{fig:twisted}
\end{figure}

How about the $\CN\ket\psi$ in the twisted sector? We thus need to consider the configuration as shown in Figure \ref{fig:twisted}, where the state after acting by $\CN$ is in the twisted sector. Since in the radial quantization, the radial direction is the time, there should be a $\Z_2$ defect line along the time/radial direction outside of the $\CN$. We again deform the $U$ operator in two ways, either shrinking inwards or expanding outwards. Shrinking inwards again yields a nontrivial sign since $\sigma$ is $\Z_2$ odd. However, by using the F-moves, expanding $U$ outwards also yields a minus sign, which comes from $F_{U,\CN, U}^{\CN}= -1$. Hence two ways of deforming $U$ does not lead to any constraint, and indeed $\CN\ket\psi$ is in general non-vanishing. Similar discussions can be applied when we insert a $\Z_2$ even local operator at the origin, and the conclusions for the untwisted and twisted sectors are exchanged.

The above discussion shows that the standard fusion rule $\CN\times U= \CN$ holds only when the state after Kramers-Wannier is in the untwisted sector.  When we work within the twisted sector, the fusion rule is modified by a minus sign.

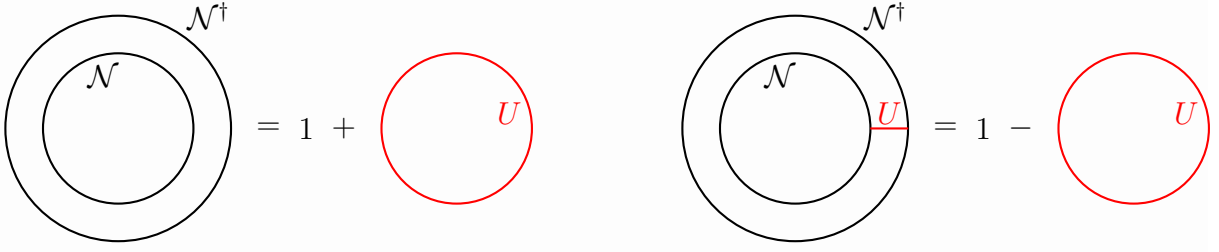
\begin{figure}
    \centering
    \begin{tikzpicture}
	\draw[thick] (0,0) circle (1);
	\node() at (-0.2,0.7){$\CN$} ;
	\draw[thick] (0,0) circle (1.5);
	\node() at (1.2,1.5){$\CN^{\dagger}$} ;
	\node() at (2,0){$=$} ;
    \node() at (2.5,0){$1$} ;
    \node() at (3,0){$+$} ;
    \draw[thick,red] (4.5,0) circle (1);
	\node[red] at (5.2,0.2){$U$} ;

	\draw[thick] (9,0) circle (1);
	\node() at (8.8,0.7){$\CN$} ;
	\draw[thick] (9,0) circle (1.5);
	\node() at (10.2,1.5){$\CN^{\dagger}$} ;
	\node() at (11,0){$=$} ;
    \node() at (11.5,0){$1$} ;
    \node() at (12,0){$-$} ;
    \draw[thick,red] (13.5,0) circle (1);
	\node[red] at (14.2,0.2){$U$} ;
    \node(b01) at (9.85,0){} ;
    \node(b02) at (10.65,0){} ;
    \path[thick,red] (b01) edge (b02);
    \node[red] at (10.25,0.2){$U$} ;
	\end{tikzpicture}
    \caption{The left is the standard fusion rule, where the state in the middle belongs to the untwisted sector. The right is the modified fusion rule, where the state in the middle belongs to the twisted sector.  }
    \label{fig:NN}
\end{figure}

We next compute the fusion rule $\CN^{\dagger}\times \CN$. To do so, we note that so far $\CN$ is only defined on the Hilbert space on sites spanned by $\ket{\{s_j\}}$, but not on the Hilbert space on links. The latter can be defined in a similar way, 
\begin{eqnarray}
    \CN^{\dagger}\ket{\{\widehat{s}_{j-\frac{1}{2}}\}} = \frac{1}{2^{L/2}} \sum_{\{s_j\}} (-1)^{\sum_{j=1}^L (\widehat{s}_{j-\frac{1}{2}}+ \widehat{s}_{j+\frac{1}{2}})s_j+ \widehat{s}_{\frac{1}{2}} t} \ket{\{s_j\}}.
\end{eqnarray}
Then $\CN^{\dagger}\times \CN$ proceeds as
\begin{equation}
\begin{split}
    \CN^{\dagger}\times \CN \ket\psi&=  \frac{1}{2^{L/2}}\sum_{\{\widehat{s}_{i-\frac{1}{2}}\},\{s_i\}} \psi_{\{s_i\}}   (-1)^{\sum_{j=1}^L (s_{j-1}+ s_{j})\widehat{s}_{j-\frac{1}{2}} + \widehat{t}s_L} \CN\ket{\{\widehat{s}_{i-\frac{1}{2}}\}}\\
    &= \frac{1}{2^{L}}\sum_{\{\widehat{s}_{i-\frac{1}{2}}\},\{s_i\}, \{s'_i\}} \psi_{\{s_i\}}   (-1)^{\sum_{j=1}^L (s_{j-1}+ s_{j})\widehat{s}_{j-\frac{1}{2}} + \widehat{t}s_L} (-1)^{\sum_{k=1}^L (\widehat{s}_{k-\frac{1}{2}}+ \widehat{s}_{k+\frac{1}{2}}) s'_k + \widehat{s}_{\frac{1}{2}} t'} \ket{\{s'_j\}}\\
    &= \frac{1}{2^{L}}\sum_{\{\widehat{s}_{i-\frac{1}{2}}\},\{s_i\}, \{s'_i\}} \psi_{\{s_i\}}   (-1)^{\sum_{j=1}^L (s_{j-1}+ s_{j}+ s'_{j-1}+ s'_{j})\widehat{s}_{j-\frac{1}{2}} + \widehat{t}(s_L+s'_L)}  \ket{\{s'_j\}}\\
    &= \sum_{\{s_i\}, \{s'_i\}}\psi_{\{s_i\}}   \delta_{s_{j-1}+ s_{j}+ s'_{j-1}+ s'_{j}}(-1)^{\widehat{t}(s_L+s'_L)}\ket{\{s'_j\}}.
\end{split}
\end{equation}
Solving the constraints for $s_{j-1}+ s_{j}+ s'_{j-1}+ s'_{j}=0$ for every $j$ yields to solutions,  $s_j'=s_j$ for all $j$, or $s_j'=s_j + 1$ for all $j$, which subsequently implies $t=t'$. In other words, 
\begin{eqnarray}
\begin{split}
    \CN^{\dagger}\times \CN \ket\psi= \sum_{\{s_i\}} \psi_{\{s_i\}}   \ket{\{s_j\}} + \sum_{\{s_i\}}  \psi_{\{1-s_i\}}  (-1)^{\widehat{t}} \ket{\{s_j\}}= (1+ (-1)^{\widehat{t}}U) \ket\psi.
\end{split}
\end{eqnarray}
This implies the fusion rule 
\begin{eqnarray}\label{eq:NNfusionrule}
\CN^{\dagger}\times \CN = 1+ (-1)^{\widehat{t}} U.
\end{eqnarray}
The fusion rule \eqref{eq:NNfusionrule} again differs from the standard one $\CN^{\dagger}\times \CN=1+U$ by a factor $(-1)^{\widehat{t}}$.
This means that the standard fusion rule~\eqref{eq:IsingFusionrule} holds only when the state in the intermediate state
(after applying one Kramers-Wannier transformation) is in the untwisted sector.
Indeed this is as expected, because if the intermediate state is in the twisted sector, there is a $U$ connecting the two $\CN$'s.  See Figure \ref{fig:NN}.

\subsection{Mapping between symmetry-twist sectors}
\label{sec:KWsectors}

We have already shown in Eq.~\eqref{eq:hatt_u} that the spatial twist (boundary condition) $\hat{t}$ of the
dual spins is linked to the spin-flip parity $u$ of the original spins, via the Kramers-Wannier transformation.

Let us also clarify the relation between the spin-flip parity $\hat{u}$ of the dual spins and the spatial twist $t$ of the original spins. We assume $\ket{\psi}$ is with in the sector labeled by $(u,t)$, i.e. 
\begin{eqnarray}
\psi_{\{1-s_i\}} = (-1)^u \psi_{\{s_i\}}, \hspace{1cm} s_{i+L} = s_i + t.
\end{eqnarray}

To determine the symmetry-twist sectors under Kramers-Wannier transformation, we first compute the resulting state using the representation~\eqref{eq:defKW_hatt} of the Kramers-Wannier transformation as
\begin{equation}\label{eq:Npsi}
\CN \ket{\psi} =  \sum_{\{\widehat{s}_{i-\frac{1}{2}}\}} \widehat{\psi}_{\{\widehat{s}_{i-\frac{1}{2}}\}}   \ket{\{\widehat{s}_{i-\frac{1}{2}}\}}, \hspace{1cm} \widehat{\psi}_{\{\widehat{s}_{i-\frac{1}{2}}\}}= \frac{1}{2^{L/2}} \sum_{\{s_i\}} \psi_{\{s_i\}} (-1)^{\sum_{j=1}^L (s_{j-1}+ s_{j})\widehat{s}_{j-\frac{1}{2}} + \widehat{t}s_L}.
\end{equation}
The symmetry sector $\widehat{u}$ is determined by $\widehat{\psi}_{\{1-\widehat{s}_{i-\frac{1}{2}}\}}= (-1)^{\widehat{u}}\widehat{\psi}_{\{\widehat{s}_{i-\frac{1}{2}}\}}$. From \eqref{eq:Npsi}, we find
\begin{eqnarray}
\begin{split}
    \widehat{\psi}_{\{1-\widehat{s}_{i-\frac{1}{2}}\}} &= \frac{1}{2^{L/2}} \sum_{\{s_i\}} \psi_{\{s_i\}} (-1)^{\sum_{j=1}^L (s_{j-1}+ s_{j})(1-\widehat{s}_{j-\frac{1}{2}}) + \widehat{t}s_L}\\
    &= \frac{1}{2^{L/2}} \sum_{\{s_i\}} \psi_{\{s_i\}} (-1)^{\sum_{j=1}^L (s_{j-1}+ s_{j})\widehat{s}_{j-\frac{1}{2}} + \widehat{t}s_L} (-1)^{\sum_{j=1}^L s_{j-1}+ s_{j}}\\
    &= (-1)^t \widehat{\psi}_{\{\widehat{s}_{i-\frac{1}{2}}\}} .
\end{split}
\end{eqnarray}
This shows that the dual $\Z_2$ symmetry after Kramers-Wannier transformation is determined by the twist before this transformation, 
\begin{eqnarray}
\label{eq:hatut}
\widehat{u} = t.
\end{eqnarray}

Combining \eqref{eq:hatut} and \eqref{eq:hatt_u}, we find that given a state $\ket{\psi}$ in the symmetry-twist sector $(u,t)$, under the Kramers-Wannier transformation, the resulting state $\CN \ket{\psi}$ is in the symmetry-twist sector $(\widehat{u},\widehat{t})=(t,u)$. The above results of the sector mapping under Kramers-Wannier transformation are summarized in Table \ref{tab:KW2d}.

\begin{table}[t]
    \centering
    \begin{minipage}{.5\linewidth}
      \centering
        \begin{tabular}{|c|cc|}
        \hline
                                    $\ket{\psi}$  & $t=0$ & $t=1$ \\
\hline
$u=0$ & \cellcolor{lightblue}  $\mS$  & \cellcolor{lightpink}  $\mU$  \\
$u=1$ & \cellcolor{lightgreen}  $\mT$  &  \cellcolor{lightyellow}  $\mV$ \\
\hline
        \end{tabular}
    \end{minipage}%
    \begin{minipage}{.5\linewidth}
      \centering
        \begin{tabular}{|c|cc|}
        \hline
            $\CN\ket\psi$  & $\widehat t=0$ & $\widehat t=1$ \\
\hline
$\widehat u=0$ & \cellcolor{lightblue}  $\mS$  &  \cellcolor{lightgreen}  $\mT$ \\
$\widehat u=1$ &  \cellcolor{lightpink}  $\mU$   & \cellcolor{lightyellow}  $\mV$ \\
\hline
        \end{tabular}
    \end{minipage} 
    \caption{Symmetry-twist sectors of the theories before and after the Kramers Wannier transformation. The $\Z_2$ even twisted sector is exchanged with the $\Z_2$ odd untwisted sector. }
    \label{tab:KW2d}
\end{table}

From the sector mapping, it is also obvious that the Kramers-Wannier transformation is not unitary, consistent with the observation in Section \ref{sec:KWnonunitary}. For instance, Kramers-Wannier transformation annihilates a $\Z_2$ odd untwisted state if one is within the untwisted sector after this transformation. (Note that we should fix one boundary condition (or twist sector) to discuss a transformation.)  Hence the probability is in general  not conserved under Kramers-Wannier transformation, and this again shows that the $\CN$ is non-unitary. The non-unitarity will further be reflected by the non-invertibility in its fusion rule.

\subsection{(Non-)unitarity of Kramers-Wannier transformation}
\label{sec:KWnonunitary}

Now let us examine the unitarity of the Kramers-Wannier transformation as defined in Eq.~\eqref{eq:defKW}.
For this purpose, we evaluate $\CN^\dagger \CN$, which should be equal to the identity operator if $\CN$ were unitary.
It can be evaluated explicitly, in the same way as the fusion rule~\eqref{eq:NNfusionrule}.
As a result, we find
\begin{eqnarray}
    \CN^\dagger \CN =  1+ (-1)^{\widehat{t}} U .
    \label{eq:CNdaggerCN}
\end{eqnarray}
The fact that the right-hand side is not identical to identity implies that $\CN$ is not
unitary for a given $\widehat{t}$.
In fact, $\CN$ annihilates the odd spin-flip parity sector ($u=1$) if $\widehat{t}=0$, or the even spin-flip parity sector ($u=0$) if $\widehat{t}=1$.
Therefore $\CN$ is not invertible.

On the other hand, we can take a different viewpoint~\cite{Seifnashri-SCGP2022}.
If we regard the spatial twists $t, \hat{t}$ as extra degrees of freedom associated to the original and dual spin systems,
or equivalently, regard the twisted and untwisted sectors as different Hilbert spaces, the Kramers-Wannier transformation
just shuffles among the different sectors as in Table \ref{tab:KW2d}
and is unitary.
Mathematically, $\widehat{t}$ can then be regarded as an extra dual spin, and thus a summation over $\widehat{t}=0,1$
should be taken in the intermediate dual spin state, resulting in $\CN^\dagger \CN=1$ from Eq~\eqref{eq:CNdaggerCN}.

\subsection{Kramers-Wannier transformation and $\Z_2$ gauging }

As mentioned earlier in this Section, the Kramers-Wannier transformation amounts to gauging the non-anomalous~\footnote{``Non-anomalous'' here refers to the nature of the symmetry
that can be gauged. The $\Z_2$ symmetry of the quantum Ising chain is non-anomalous because it is on-site; see the beginning of this Section.}
$\Z_2$ global symmetry. We start with the partition function of theory $\CX$ with a non-anomalous $\Z_2$ global symmetry whose background field is $A$, i.e. $Z_{\CX}[X_2, A]$. Gauging $\Z_2$ yields another theory $\CX/\Z_2$, whose partition function is 
\begin{eqnarray}
Z_{\CX/\Z_2}[X_2, \widehat{A}] = \frac{1}{|H^0(X_2, \Z_2)|} \sum_{a\in H^1(X_2, \Z_2)} Z_{\CX}[X_2, a] (-1)^{\int_{X_2} a \widehat{A}}
\end{eqnarray}
where $\widehat{A}$ is the background gauge field for the dual $\Z_2$ symmetry of $\CX/\Z_2$.

The mapping between the symmetry and twist sectors has been discussed in \cite{Hsieh:2020uwb,Fukusumi:2021zme, Cao:2022lig}. Let us briefly review the results. We first place the system on a torus, $X_2 = T^2$. The gauge fields can thus be replaced by  their holonomies $A\to \{W_t, W_x\}$, 
\begin{eqnarray}\label{eq:Zwtwx}
Z_{\CX/\Z_2}[\widehat{W}_t, \widehat{W}_x] = \frac{1}{2} \sum_{w_t, w_x=0,1} Z_{\CX}[ w_t, w_x] (-1)^{w_t \widehat{W}_x + w_x \widehat{W}_t}.
\end{eqnarray}
The partition functions in different symmetry and twist sectors labeled by $(u,t)$  are given by 
\begin{eqnarray}\label{eq:Zut}
\begin{split}
    Z_{\CX}^{(u,t)} = \frac{1}{2} \sum_{w_t=0,1} Z_{\CX}[w_t, t] (-1)^{u w_t}
\end{split}
\end{eqnarray}
and the converse relation is
\begin{eqnarray}\label{eq:Zwtwx2}
Z_{\CX}[w_t,w_x] = \sum_{u=0,1}Z_{\CX}^{(u,w_x)} (-1)^{u w_t}.
\end{eqnarray}
The symmetry and twist sectors for $\CX/\Z_2$ are likewisely defined. Combining \eqref{eq:Zwtwx}, \eqref{eq:Zut} and \eqref{eq:Zwtwx2}, we find
\begin{eqnarray}
\begin{split}
    Z_{\CX/\Z_2}^{(\widehat{u}, \widehat{t})} &= \frac{1}{2} \sum_{\widehat{w}_t=0,1} Z_{\CX/\Z_2}[\widehat{w}_t, \widehat{t}] (-1)^{\widehat{u} \widehat{w}_t}\\
    &= \frac{1}{4} \sum_{\widehat{w}_t=0,1} \sum_{w_t,w_x=0,1} Z_{\CX}[{w}_t, w_x]  (-1)^{w_t \widehat{t} + w_x \widehat{w}_t + \widehat{u} \widehat{w}_t}\\
    &= \frac{1}{4} \sum_{\widehat{w}_t=0,1} \sum_{w_t,w_x=0,1} \sum_{u=0,1} Z_{\CX}^{(u,w_x)}  (-1)^{u w_t+ w_t \widehat{t} + w_x \widehat{w}_t + \widehat{u} \widehat{w}_t}\\
    &=  \sum_{w_x=0,1} \sum_{u=0,1} Z_{\CX}^{(u,w_x)}   \delta_{u,\widehat{t}} \delta_{w_x, \widehat{u}}= Z_{\CX}^{(\widehat{t}, \widehat{u})}.
\end{split}
\end{eqnarray}
This is precisely the mapping between symmetry and twist sectors $(\widehat{u},\widehat{t}) = (t,u)$ derived using the Kramers-Wannier transformation on the lattice.

The fusion rule of the topological interface between $\CX$ and $\CX/\Z_2$ can also be derived, following \cite{Kaidi:2021xfk, Choi:2021kmx, Choi:2022zal,Kaidi:2022cpf}. 
We will not repeat the derivation here, and refer the interested readers to these references, e.g. Section 2 of \cite{Kaidi:2022cpf}. One remark is that in deriving the fusion rule between the duality interfaces $\CN^{\dagger}\times \CN$, one does not turn on the $\Z_2$ defects $U$ in the vicinity of the locus of $\CN$, hence the fusion rule corresponds to  the left panel of Figure \ref{fig:NN}. This point has already been emphasized in \cite{Choi:2022zal}.

\subsection{Kramers-Wannier transformation on the  transverse field Ising Hamiltonian}

The Hamiltonian is a sum over local interactions given by the Pauli operators. Let us first consider how the Pauli operators are mapped under Kramers-Wannier transformation. It is straightforward to check that
\begin{eqnarray}\label{eq:Isingmap}
\tau^x_{i-\frac{1}{2}} \CN \ket{\psi} = \CN \sigma^z_{i-1} \sigma^z_{i} \ket\psi, \hspace{1cm} \tau^z_{i-\frac{1}{2}} \tau^z_{i+\frac{1}{2}} \CN \ket{\psi} = \CN  \sigma^x_{i} \ket\psi
\end{eqnarray}
where $i=1, ..., L$. 
{Now let us consider the transverse Ising chain~\eqref{eq:IsingH}.
When the system is defined on a ring, the Hamiltonian is more precisely specified as
\begin{equation}
    H_{\text{Ising}}^h = - \sum_{i=1}^L \left( \sigma^z_{i-1}\sigma^z_{i} + h \sigma^x_{i}\right),
\label{eq:IsingH_ring}
\end{equation}
with the identification of site $0$ with site $L$ as in Eq.~\eqref{eq:Ising_bc}.
Note that the boundary conditions are already encoded into the Hilbert spaces.
For example, $\sigma^z_0\ket{s_0}= (-1)^{s_0} \ket{s_0}= (-1)^{s_L+t} \ket{s_0}$. Hence effectively $\sigma^z_0= (-1)^t \sigma^z_L$, and one should replace the term $\sigma^z_{0} \sigma^z_1$ by $(-1)^L\sigma^z_{L} \sigma^z_1$, which is a more common convention used in the literature (for example \cite{yao2021twisted,Li:2022drc}).
Using the above map \eqref{eq:Isingmap}, the Kramers-Wannier dual Hamiltonian of the transverse Ising chain~\eqref{eq:IsingH_ring} is 
\begin{eqnarray}
\widehat{H}_{\text{Ising}}^{h} = - \sum_{i=1}^{L} \left( \tau^x_{i-\frac{1}{2}} + h \tau^z_{i-\frac{1}{2}} \tau^z_{i+\frac{1}{2}}\right).
\label{eq:dualIsingH}
\end{eqnarray}
By shifting the spins on the links to the sites (which is simply a relabeling), one finds that 
\begin{eqnarray}
H^h_{\text{Ising}} = h \widehat{H}^{1/h}_{\text{Ising}}.
\end{eqnarray}}


Given the duality mapping of the Hamiltonian, we can see why the Kramers-Wannier transformation
\emph{must} be non-invertible, and why the unitarity can be restored by expanding the Hilbert space
by including the twisted sector.
Let us consider the ordered phase $h \ll 1$ of the Ising model of the original spins.
The ground states are two-fold degenerate, corresponding to the spontaneous magnetization ``up'' and ``down''.
The Kramers-Wannier transformation maps this model to the Ising model of the dual spins in the disordered phase,
where the ground state is unique.
As a consequence, the Kramers-Wannier transformation must map the two ground states to one, and thus is non-unitary and non-invertible.
In our construction, the 2-to-1 mapping is achieved by projecting out one of the spin-flip parity sectors.

On the other hand, the Ising model in the disordered phase is insensitive to the boundary condition.
Therefore, the ground-state energy under the twisted boundary condition is asymptotically degenerate
with that under the periodic boundary condition.
If we expand the Hilbert space by including the twisted sector, the ground states are two-fold degenerate.
In contrast, when the Ising model is in the ordered phase, the twisted boundary condition
introduces a domain wall with a non-zero energy. As a consequence, the ground-state energy in the
twisted sector is higher than that in the untwisted sector (periodic boundary condition).
Thus the ground-state degeneracy remains 2 (coming from the spontaneous breaking of the spin-flip symmetry)
even if the Hilbert space is expanded.
The Kramers-Wannier transformation between the extended Hilbert spaces can be
invertible (and unitary).
Of course the present argument which focuses on the ground state alone does not prove the
invertibility or unitarity, but it gives a physical perspective on the unitarity we have shown
by an explicit calculation in Sec.~\ref{sec:KWnonunitary}.

\subsection{Kramers-Wannier transformation on an interval: A unitary transformation}
\label{sec:KWopen}

We proceed to discuss the Kramers-Wannier transformation on an open chain.
Although the operator $\CN$ implementing the Kramers-Wannier transformation is non-unitary and satisfies the non-invertible fusion rule,
the $\CN$ under certain open boundary conditions is unitary~\cite{Nussinov:2012nz}.

Suppose the open chain contains sites $i=1, ..., L$, and the dual spins live on half-integer links $i-\frac{1}{2}$ for $i=1, ..., L$. 
We begin by modifying \eqref{eq:defKW} such that only the terms that are fully supported will be kept in the exponent, i.e. free boundary condition. Concretely, 
\begin{eqnarray}
\CN \ket{\{s_i\}} = \frac{1}{2^{L/2}} \sum_{\{\widehat{s}_{i-\frac{1}{2}}\}} (-1)^{\sum_{j=2}^L s_{j-1}\widehat{s}_{j-\frac{1}{2}} + \sum_{j=1}^L s_{j}\widehat{s}_{j-\frac{1}{2}} } \ket{\{\widehat{s}_{i-\frac{1}{2}}\}}.
\end{eqnarray}
Note that we also dropped the term $\widehat{t} S_{L}$ because the twisted boundary condition is well defined only on closed chains. 

It is immediate to check that $\CN$ is a unitary transformation, by directly checking $\bra{\{ s_i\}} \CN^\dagger \CN \ket{\{ s'_i\}}$. To see this, we compute 
\begin{equation}
\begin{split}
    &\bra{\{ s_i\}} \CN^\dagger \CN \ket{\{ s'_i\}}\\&= \frac{1}{2^L} \sum_{\{\widehat{s}_{i-\frac{1}{2}}\}, \{\widehat{s}'_{i-\frac{1}{2}}\}  }  \bra{\{\widehat{s}_{i-\frac{1}{2}}\}}   (-1)^{\sum_{j=2}^L s_{j-1}\widehat{s}_{j-\frac{1}{2}} + \sum_{j=1}^L s_{j}\widehat{s}_{j-\frac{1}{2}} } (-1)^{\sum_{j=2}^L s'_{j-1}\widehat{s}'_{j-\frac{1}{2}} + \sum_{j=1}^L s'_{j}\widehat{s}'_{j-\frac{1}{2}} } \ket{\{\widehat{s}'_{i-\frac{1}{2}}\}}\\
    &= \frac{1}{2^L} \sum_{\{\widehat{s}_{i-\frac{1}{2}}\} }   (-1)^{\sum_{j=2}^L (s_{j-1}+ s'_{j-1} )\widehat{s}_{j-\frac{1}{2}} + \sum_{j=1}^L (s_{j}+ s'_{j} )\widehat{s}_{j-\frac{1}{2}}} 
    =    \prod_{i=1}^{L}\delta_{s_i, s_i'}  .
\end{split}
\end{equation}
This shows that $\CN^\dagger \CN=I$, hence $\CN$ is a unitary operator. It should be contrasted to the non-unitarity of $\CN$ on the closed chain.

It is useful to find the mapping between Pauli operators. Our goal is to solve $\CO^z_j(\{\sigma^{x,z}_k\})$ satisfying
\begin{eqnarray}
\tau^x_{j-\frac{1}{2}} \CN \ket\psi = \CN \CO^{x}_{j}(\{\sigma^{x,z}_k\}) \ket\psi, \hspace{1cm}\tau^z_{j-\frac{1}{2}} \CN \ket\psi = \CN \CO^{z}_{j}(\{\sigma^{x,z}_k\}) \ket\psi
\end{eqnarray}
for any $\ket\psi$. 
The calculation is straightforward, and the result is
\begin{eqnarray}
\begin{split}
     \CO^{x}_{j}(\{\sigma^{x,z}_k\}) = 
     \begin{cases}
     \sigma_{j-1}^z \sigma_j^z, & j=2, ..., L\\
     \sigma_1^z, & j=1
     \end{cases}, \hspace{1cm}
     \CO^{z}_{j}(\{\sigma^{x,z}_k\}) = \prod_{k=j}^{L} \sigma_k^x.
     \label{eq:tauz_sigmax}
\end{split}
\end{eqnarray}
As a consistency check, the commutation relations between $\tau^{x,z}_{j-\frac{1}{2}}$ match those between $\CO^{x,z}_{j}(\{\sigma^{x,z}_k\})$. These maps will become useful in Section \ref{sec:projtospin1}.

Let us now discuss the Kramers-Wannier transformation of the Ising model Hamiltonian on an open chain, which was also discussed in \cite{cobanera2011bond}.
Because of the mapping, the standard Ising model defined on the open chain is not
exactly self-dual; Eqs.~\eqref{eq:IsingH} is \emph{not} mapped to Eq.~\eqref{eq:dualIsingH} by the Kramers-Wannier transformation on the open chain.
More precisely, we find 
\begin{eqnarray}
H_{\text{open Ising}}^h = - \sigma^z_1 - \sum_{i=2}^{L} \sigma^z_{i-1}\sigma^z_{i} - h\sum_{i=1}^L  \sigma^x_{i} 
\label{eq:openIsingH}
\end{eqnarray}
is dual to 
\begin{eqnarray}
\widehat{H}_{\text{open Ising}}^{h} =  - \sum_{i=1}^{L} \tau^x_{i-\frac{1}{2}} - h \sum_{i=1}^{L-1} \tau^z_{i-\frac{1}{2}} \tau^z_{i+\frac{1}{2}} 
 - h \tau^z_{L-\frac{1}{2}} .
\end{eqnarray}
Note the existence of the longitudinal magnetic fields (coupled to the $z$-component of the spin) at the boundary
in either side.
The boundary longitudinal magnetic field breaks the spin-flip symmetry explicitly.
As a consequence, the ground state is unique (chosen by the boundary longitudinal magnetic field) even in the ordered phase.
This resolves the obstacle to the unitarity of the Kramers-Wannier transformation discussed in Sec.~\ref{sec:KWnonunitary}, as
the mapping of the ground states now becomes 1 to 1.

Alternatively, to maintain the spin-flip symmetry of the original spins, we can omit the boundary longitudinal field $-\sigma^z_1$
in Eq.~\eqref{eq:openIsingH}.
In this case, the ground state in the ordered phase $h<1$ is two-fold degenerate, reflecting the spontaneous symmetry breaking.
The dual Hamiltonian lacks the transverse field $- \tau^x_{\frac{1}{2}}$ at the end of the chain.
The edge spin at $1/2$ is still coupled to the neighboring one by the Ising coupling $\tau^z_{\frac{1}{2}} \tau^z_{\frac{3}{2}}$.
Therefore, in the ordered phase of the dual spins (which corresponds to the disordered phase of the original spins), the end
spin is polarized and the ground state is unique (since the spin-flip symmetry is explicitly broken by the boundary longitudinal field at the other
end $L-\frac{1}{2}$).
On the other hand, in the disordered phase of the dual spins (which corresponds to the ordered phase of the original spins),
the end spin at $1/2$ has no favored direction; this implies the presence of the spin-$1/2$ ``edge mode'' and the ground states
are two-fold degenerate.
As a result, the Kramers-Wannier transformation does preserve the number of the ground states: 1 to 1 mapping in the disordered phase (of the original spins)
and 2 to 2 in the ordered phase. This means that Kramers-Wannier transformation can be unitary, as it was indeed shown by the explicit calculations.

The fact that the spontaneous symmetry breaking of the global symmetry of the original spin system
corresponds to the edge state in the dual spin system can be also read off from the
mapping~\eqref{eq:tauz_sigmax}. The global spin-flip operator $U = \prod_{j=1}^L \sigma_j^x$ is mapped to the single spin
operator $\tau^z_{\frac{1}{2}}$ at the end of the dual spin chain.

The Kramers-Wannier duality between the spontaneous breaking of the global symmetry and the edge state is reminiscent of the
Kennedy-Tasaki transformation.
However, there are important differences. While the Kramers-Wannier transformation on an open chain maps the global symmetry generator to the local operator at the end of the
chain as we have seen above,
the Kennedy-Tasaki transformation on an open chain preserves the global symmetry generators as in Eq.~\eqref{eq:R_UKT}.
Nevertheless, they are deeply related, as we will see in the following sections.

\section{Kennedy-Tasaki transformation on a ring of spin-1 per unit cell}
\label{sec:KTspin1ring}

As we have discussed in the Introduction, the Kennedy-Tasaki transformation as a unitary transformation has been
discussed exclusively for open boundary conditions.
The transformation~\eqref{eq:KToriginal} appears to be ill-defined for the periodic boundary conditions, as it depends on the
ordering of the sites. Furthermore, the SPT phase with a unique ground state should be mapped to the phase breaking
the $\Z_2 \times \Z_2$ symmetry with 4 degenerate ground states, which seems impossible with a unitary transformation.
However, as we have seen in Secs.~\ref{sec:KWsectors} and~\ref{sec:KWnonunitary}, the latter problem could be
resolved by matching the symmetry sector and the boundary condition in the case of Kramers-Wannier transformation.

To formulate the Kennedy-Tasaki transformation \eqref{eq:KToriginal} on a ring, we have to overcome two apparent difficulties mentioned in the introduction: 1) the lack of natural ordering on a ring, and 2) the mismatch of ground state degeneracy. The second difficulty has been briefly mentioned above, and we have seen a resolution in the case of the Kramers-Wannier transformation. We will come back to this later. 

How to address the ordering problem on a ring? The key observation is that although it seems hard to define the unitary operator $U_{\text{KT}}$ on a ring, the transformation on spin-1 operators \eqref{eq:Sx_KT}, \eqref{eq:Sz_KT} and \eqref{eq:Sy_KT} have a natural definition on the ring! Note that the Kennedy-Tasaki transformation simply addresses a string operator that generates the $\Z_2^x\times \Z_2^z$ symmetry to the spins, similar to both Kramers-Wannier and Jordan-Wigner transformations \cite{Cao:2022lig,Li:2022jbf, RevModPhys.36.856,Hsieh:2020uwb}.  We thus \emph{define} the Kennedy-Tasaki transformation on a ring by specifying how the spin-1 operators map:
\begin{eqnarray}\label{eq:Sx_KT1}
\begin{split}
    S'^{x}_{j} &= S^x_j e^{i \pi \sum_{k=1}^{j-1} S^x_k},
\\
S'^{z}_{j} &= e^{i\pi \sum_{k=j+1}^{L} S^z_k} S^z_j =  R^z e^{i\pi \sum_{k=1}^{j} S^z_k} S_j^z ,
\\
S'^{y}_{j} &= e^{i\pi \sum_{k=j+1}^{L} S^z_k} S^y_j e^{i \pi \sum_{k=1}^{j-1} S^x_k } .
\end{split}
\end{eqnarray}
where $j=1, ..., L$. When $j=1$, the string $e^{i \pi \sum_{k=1}^{j-1} S^x_k}=1$ is trivial; when $j=L$, the string $e^{i\pi \sum_{k=j+1}^{L} S^z_k}=1$ is trivial. Indeed, the commutation relations among $S'^{x,y,z}_{j}$ are still those of the standard spin-1 operators. 

\paragraph{Mapping between symmetry sectors:}
We proceed to discuss how the symmetry sectors and boundary conditions transform under the map \eqref{eq:Sx_KT1}. First symmetry operators $R^{\alpha}= e^{i\pi \sum_{j=1}^{L} S^\alpha_j}$ are invariant under the transformation, 
\begin{eqnarray}
    R'^{\alpha} = R^{\alpha}, \hspace{1cm} \alpha=x,y,z ,
\end{eqnarray}
as we have shown in Eq.~\eqref{eq:R_UKT}. Denoting the eigenvalue of $R^{\alpha}$ as $(-1)^{u_\alpha}$, we thus have
\begin{eqnarray}\label{eq:ux}
    u'_x=u_x \mod 2, \hspace{1cm} u'_z=u_z \mod 2.
\end{eqnarray}

\paragraph{Mapping between boundary conditions:}
The boundary condition is specified by 
\begin{eqnarray}
    S_{j+L}^x= (-1)^{t_z} S_j^x, \hspace{1cm} S_{j+L}^z= (-1)^{t_x} S_j^z.
\end{eqnarray}
Note that under $\Z_2^x$ generated by $R^x$, the $S_j^z$ flips sign, hence the boundary condition for $S_j^z$ is labeled by $t_x$. Then the boundary condition for the $S'^{x,z}_j$ can be determined from \eqref{eq:Sx_KT1} via 
\begin{eqnarray}
\begin{split}
    S'^x_{j+L}&= S^x_{j+L} e^{i \pi \sum_{k=1}^{j+L-1} S^x_k} = (-1)^{t_z} S^x_j e^{i \pi \sum_{k=1}^{j-1} S^x_k} R^x = (-1)^{t_z+ u_x} S^x_{j},\\
    S'^z_{j+L}&= R^z S^z_{j+L} e^{i \pi \sum_{k=1}^{j+L} S^x_k} = (-1)^{t_x} R^z S^z_j e^{i \pi \sum_{k=1}^{j} S^z_k} R^z = (-1)^{t_x+ u_z} S^z_{j}\\
\end{split}
\end{eqnarray}
which implies
\begin{eqnarray}\label{eq:tx}
    t'_z= t_z+u_x \mod 2, \hspace{1cm} t'_x= t_x+ u_z \mod 2.
\end{eqnarray}
Although the generalization of the Kennedy-Tasaki transformation from the open chain to the closed chain as in \eqref{eq:Sx_KT1} looks naive, the mapping between the symmetry and twist sectors \eqref{eq:ux} and \eqref{eq:tx} match precisely with mapping induced by the twisted gauging to be discussed in Section \ref{sec:KTsectormap}. The above relations are also very much reminiscent of the
similar, well-known relation for the Jordan-Wigner transformation on a ring~\cite{RevModPhys.36.856,Hsieh:2020uwb}.

In condensed matter literatures, the boundary condition is more conventionally specified by modifying one term in the Hamiltonian that crosses the boundary. 
It is useful to derive \eqref{eq:tx} from this conventional point of view. For example, let's consider the Heisenberg Hamiltonian under $\Z_2^x\times \Z_2^z$ twisted boundary condition
\begin{equation}\label{eq:HambeforeKTspin1}
    H=  \sum_{i=1}^{L-1} \left(J_x S^x_j S^x_{j+1} +J_y S^y_j S^y_{j+1} + J_z S^z_j S^z_{j+1}\right) + (-1)^{t_z} J_x S^x_L S^x_1 + (-1)^{t_z+t_x} J_y S^y_L S^y_1 + (-1)^{t_x} J_z S^z_L S^z_1.
\end{equation}
Under Kennedy-Tasaki transformation, the above Hamiltonian is mapped to 
\begin{equation}\label{eq:HamafterKTspin1}
\begin{split}
    H_{\text{KT}} =&  \sum_{i=1}^{L-1} \left( J_xe^{i\pi S_j^x} S^x_j S^x_{j+1} + J_ye^{i\pi S_j^x} S^y_j S^y_{j+1} e^{i\pi S_{j+1}^z} + J_z S^z_j S^z_{j+1} e^{i\pi S_{j+1}^z}  \right)\\
    &+ (-1)^{t_z+u_x} J_x e^{i\pi S_L^x} S^x_L S^x_1 + (-1)^{t_z+t_x+u_x+u_z} J_y e^{i\pi S_L^x} S^y_L S^y_1 e^{i\pi S_1^z} + (-1)^{t_x+u_z} J_z  S^z_L S^z_1 e^{i\pi S_1^z}
\end{split}
\end{equation}
from which we again read off the mapping between the boundary conditions and the symmetry sectors as in  \eqref{eq:ux} and \eqref{eq:tx}.

This also resolves the issue of the ground-state degeneracy for the Kennedy-Tasaki transformation on the ring.
Similarly to the discussion on the Kramers-Wannier transformation in Sec.~\ref{sec:KWnonunitary},
spontaneous breaking of the $\Z_2 \times \Z_2$ symmetry implies a 4-fold ground state degeneracy
under the periodic boundary condition, with one ground state in each of the 4 symmetry sectors
$u_z, u_x = 0,1$. Since a twisted boundary condition introduces a domain wall with positive energy,
the twisted sectors $t_z=1$ or $t_x=1$ do not contribute ground states in the extended Hilbert space. On the other hand, the Kennedy-Tasaki dual of the symmetry-broken phase is the Haldane SPT phase, which
does not have a long-range order and thus is insensitive to the boundary conditions.
As a consequence, the ground states in each of the 4 twist sectors $t'_z, t'_x =0,1$
are degenerate, resulting in the 4-fold ground-state degeneracy (in the extended Hilbert space).
Therefore, \emph{in the extended Hilbert space}, the Kennedy-Tasaki transformation induces a 4-to-4 mapping of the ground states
and thus can be unitary.
This is analogous to the unitarity of the Kramers-Wannier mapping in the extended Hilbert space,
as discussed in Sec.~\ref{sec:KWnonunitary} and in Ref.~\cite{Seifnashri-SCGP2022}.
If we focus on the untwisted Hilbert space only (i.e. PBC), the Kennedy-Tasaki transformation is non-unitary.

\section{Field-theory formulation of the Kennedy-Tasaki transformation}
\label{sec:fieldtheory}

In the previous section, we have observed that the Kennedy-Tasaki transformation can be defined on a ring by sacrificing  unitarity. Moreover, the unitarity can be restored by including twisted sectors.
In order to obtain deeper insights, let us formulate the Kennedy-Tasaki transformation in terms of field theory.

Let us denote the partition function of an arbitrary QFT $\CX$ with non-anomalous $\Z_2\times \Z_2$ symmetry as $Z_{\CX}[A_1, A_2]$, where $A_i$ is the background field for the $i$-th $\Z_2$. When $\CX$ is in the trivial phase, the fixed point partition function is
\begin{eqnarray}\label{eq:tri}
Z_{\text{Tri}}[A_1, A_2]= 1.
\end{eqnarray}
When $\CX$ is in the $\Z_2\times \Z_2$ SSB phase, the fixed point partition function is 
\begin{eqnarray}\label{eq:ssb}
Z_{\text{SSB}}[A_1, A_2]= \delta(A_1) \delta(A_2).
\end{eqnarray}
When $\CX$ is in the $\Z_2\times \Z_2$ SPT phase, the fixed point partition function is \cite{PhysRevLett.114.031601,PhysRevB.91.035134}
\begin{eqnarray}\label{eq:spt}
Z_{\text{SPT}}[A_1, A_2]=(-1)^{\int A_1 A_2}.
\end{eqnarray}
To see how these theories are related, we define the following topological manipulations: 
\begin{eqnarray}
\begin{split}
    S: Z_{S_{12}\CX}[A_1, A_2]&:= \frac{1}{|H^0(X_2, \Z_2)|^2} \sum_{a_1, a_2\in H^1(X_2, \Z_2)} Z_{\CX}[a_1, a_2] (-1)^{\int_{X_2} a_1 A_2 + a_2 A_1}  \\
    T: Z_{T_{12}\CX}[A_1, A_2]&:= Z_{\CX}[A_1, A_2] (-1)^{\int_{X_2} A_1 A_2}.
\end{split}
=' 
\end{eqnarray}
The first topological manipulation $S$ is gauging the $\Z_2\times \Z_2$. \footnote{It is also possible to discuss one of the two $\Z_2$'s. However, for our purpose we will not consider them in this work. } The second one is stacking a $\Z_2\times \Z_2$ SPT. With the above operations, we are able to fit the three theories \eqref{eq:tri}, \eqref{eq:ssb} and \eqref{eq:spt} into the following web, 
\begin{eqnarray}
\begin{tikzpicture}
[baseline=0,scale = 0.7, baseline=-10]
 \node[below] (1) at (0,0) {SSB};
  \node[below] (2) at (6,0) {Trivial};
   \node[below] (3) at (12,0) {SPT};
   
       \draw [thick,{Latex[length=2.5mm]}-{Latex[length=2.5mm]}] (1) -- (2) node[midway,  above] {$S$};
        \draw [thick,{Latex[length=2.5mm]}-{Latex[length=2.5mm]}] (2) -- (3) node[midway,  above] {$T$};
       
\draw[thick,{Latex[length=2.5mm]}-{Latex[length=2.5mm]}] (1) to[out=135, in=225,loop] (1);
  \node[left]  at (-2.5,-0.5) {$T$};
  
\draw[thick,{Latex[length=2.5mm]}-{Latex[length=2.5mm]}] (3) to[out=45, in=-45,loop] (3);
  \node[right]  at (14.5,-0.5) {$S$};
\end{tikzpicture} 
\end{eqnarray}
The only combination of the topological manipulations that exchanges SPT and SSB while preserving the trivial phase is 
\begin{eqnarray}
STS = TST.
\end{eqnarray}
The two expressions are related by the identity of $SL(2,\Z_2)$, i.e. $(ST)^3=1$. However, it will become clear that when formulating the topological manipulations on an open chain, there are subtle differences between $STS$ and $TST$, and $STS$ turns out to be simpler which is what we will use. 
The above discussion strongly suggests that the Kennedy-Tasaki transformation should simply be the $STS$ transformation.

In the following sections, we will implement $STS$ transformation, which was conceived in field theory, on spin chains.
For this purpose, it is convenient to consider spin chains with two spin-$\frac{1}{2}$'s per unit cell, rather than the spin-1 chains discussed in the original literature on
the Kennedy-Tasaki transformation. The implementation on the spin-$\frac{1}{2}$ models is also useful in elucidating the deep connection between the Kramers-Wannier and
Kennedy-Tasaki transformations.
We will also discuss the relation between the Kennedy-Tasaki transformation for the spin-$\frac{1}{2}$ models and the original Kennedy-Tasaki transformation for the spin-1 chains.

\section{Kennedy-Tasaki transformation on a ring of two spin-$\frac{1}{2}$'s per unit cell}
\label{sec:KTclosedbdy}

In this section, we discuss the  Kennedy-Tasaki transformation implementing $STS$ on a ring. Parallel to the study of Kramers-Wannier transformation, we study the fusion rules, mapping between symmetry and twist sectors, mapping between local operators, etc.

\subsection{Non-invertible Kennedy-Tasaki transformation}

Let us consider a spin chain with $L$ sites and $L$ links. Each site supports one spin-$\frac{1}{2}$, spanning a two dimensional local Hilbert space $\ket{s^{\sigma}_i}$, where $s^{\sigma}_i=0,1$ and $i=1, ..., L$. Moreover, each link also supports one spin-$\frac{1}{2}$ spanning a two dimensional local Hilbert space $\ket{s^\tau_{i-\frac{1}{2}}}$, where $s^\tau_{i-\frac{1}{2}}=0,1$ for $i=1, ..., L$. Hence each unit cell contains two spin-$\frac{1}{2}$'s.\footnote{This should be distinguished from the situation in Section \ref{sec:Kramers-Wannier} where each unit cell only supports one spin-$\frac{1}{2}$, and they either live on sites or on links, but not both. } The local states can be acted upon by Pauli operators, 
\begin{eqnarray}
\begin{split}
    \sigma_i^z\ket{s^{\sigma}_i}= (-1)^{s^{\sigma}_i}\ket{\ssi_i}, \hspace{1cm}& \sigma^x_i\ket{\ssi_i}= \ket{1-\ssi_i}\\
    \tau^z_{i-\frac{1}{2}} \ket{\st_{i-\frac{1}{2}}} = (-1)^{\st_{i-\frac{1}{2}}}\ket{\st_{i-\frac{1}{2}}}, \hspace{1cm}& \tau^x_{i-\frac{1}{2}} \ket{\st_{i-\frac{1}{2}}}= \ket{1-\st_{i-\frac{1}{2}}}.
\end{split}
\end{eqnarray}
The $\Z_2\times \Z_2$ symmetry is generated by $U_{\sigma}$ and $U_{\tau}$ respectively, where
\begin{eqnarray}\label{eq:Z2Z2op}
U_{\sigma}= \prod_{i=1}^L \sigma^x_{i}, \hspace{1cm} U_{\tau}= \prod_{i=1}^L \tau^x_{i-\frac{1}{2}}.
\end{eqnarray}
The symmetry and twist sectors are labeled by $(u_\sigma, u_\tau, t_\sigma, t_\tau)$. Here $u_\sigma, u_\tau$ are the eigenvalues of $U_\sigma, U_\tau$ respectively, and $t_\sigma, t_\tau$ label the boundary conditions $\ssi_{i+L}= \ssi_i+t_\sigma, \st_{i-\frac{1}{2}+L}= \st_{i-\frac{1}{2}}+t_\tau$.

After the Kramers-Wannier transformation, i.e. $S$ transformation which gauges $\Z_2\times \Z_2$, the spins on sites and the spins on links are exchanged, and we denote the resulting spins as $\hssi_{i-\frac{1}{2}}$ and $\hst_{i}$. Likewise, the dual spins can also be organized into 16 symmetry and twist sectors as $(\widehat{u}_\sigma, \widehat{u}_\tau, \widehat{t}_\sigma, \widehat{t}_\tau)$. Following the definition \eqref{eq:defKW}, the Kramers-Wannier transformation for both $\Z_2$'s is 
\begin{eqnarray}
\CN\ket{\{\ssi_i, \st_{i-\frac{1}{2}} \}} = \frac{1}{2^L} \sum_{\{\hssi_{j-\frac{1}{2}}, \hst_{j} \}} (-1)^{\sum_{j=1}^L \ssi_j(\hssi_{j-\frac{1}{2}}+ \hssi_{j+\frac{1}{2}}) + t_\sigma \hssi_{\frac{1}{2}} + \hst_j(\st_{j-\frac{1}{2}}+ \st_{j+\frac{1}{2}}) + \widehat{t}_\tau \st_{\frac{1}{2}} } \ket{\{\hssi_{j-\frac{1}{2}}, \hst_{j}\}}.
\end{eqnarray}
The $T$ transformation amounts to stacking a $\Z_2\times \Z_2$ SPT, and the operator implementing such stacking has been discussed in \cite{scaffidi2017gapless, Li:2022jbf, Li:2022nwa}, under the name of domain wall decoration $U_{\text{DW}}$. The $U_{\text{DW}}$ acts on the basis state as
\begin{eqnarray}
U_{\text{DW}} \ket{\{\hssi_{i-\frac{1}{2}}, \hst_{i} \}} = (-1)^{\sum_{j=1}^L \hst_j(\hssi_{j-\frac{1}{2}} + \hssi_{j+\frac{1}{2}}) + \widehat{t}_\tau \hssi_{\frac{1}{2}} } \ket{\{\hssi_{i-\frac{1}{2}}, \hst_{i} \}}.
\end{eqnarray}
This is a unitary operator. 
In the above we only defined how $U_{\text{DW}}$ acts on the dual spins, but the action on the original spins can also be similarly defined. By definition, the $STS$ transformation is defined to be the product $\CN_{\text{KT}}= \CN^{\dagger} U_{\text{DW}} \CN$. When acting on an arbitrary basis state, we find the  Kennedy-Tasaki transformation , 
\begin{eqnarray}\label{eq:ktdef2}
\begin{split}
    \CN_{\text{KT}} \ket{\{\ssi_i, \st_{i-\frac{1}{2}} \}} &= \frac{1}{2^{L+1}} \sum_{\{\sprimesi_{i}, \sprimet_{i-\frac{1}{2}}\}}(-1)^{\sum_{j=1}^L (\ssi_j + \sprimesi_j)(\st_{j-\frac{1}{2}} + \st_{j+\frac{1}{2}} + \sprimet_{j-\frac{1}{2}}+ \sprimet_{j+\frac{1}{2}}) + (\st_{\frac{1}{2}}+ \sprimet_{\frac{1}{2}})(t_\sigma + t'_{\sigma})}\\
    &\hspace{3cm}\left( 1+ (-1)^{t_\sigma + t'_{\sigma}+ \widehat{t}_{\tau}}\right) \left( 1+ (-1)^{{t}_\tau + t'_{\tau}+ \widehat{t}_{\sigma}}\right) \ket{\{\sprimesi_i, \sprimet_{i-\frac{1}{2}} \}}
\end{split}
\end{eqnarray}
where $\widehat{t}_{\tau}, \widehat{t}_{\sigma}$ label the twist sectors in the intermediate state after one Kramers-Wannier transformation.
Since they only appear in the projectors in the second line, it means that the twist sectors in the intermediate state are completely determined by the twist sectors of the initial and final states, hence \eqref{eq:ktdef2} simplifies to 
\begin{equation}\label{eq:ktdef}
\begin{split}
    \CN_{\text{KT}} \ket{\{\ssi_i, \st_{i-\frac{1}{2}} \}} &= \frac{1}{2^{L-1}} \sum_{\{\sprimesi_{i}, \sprimet_{i-\frac{1}{2}}\}}(-1)^{\sum_{j=1}^L (\ssi_j + \sprimesi_j)(\st_{j-\frac{1}{2}} + \st_{j+\frac{1}{2}} + \sprimet_{j-\frac{1}{2}}+ \sprimet_{j+\frac{1}{2}}) + (\st_{\frac{1}{2}}+ \sprimet_{\frac{1}{2}})(t_\sigma + t'_{\sigma})}\ket{\{\sprimesi_i, \sprimet_{i-\frac{1}{2}} \}}.
\end{split}
\end{equation}

\subsection{Mapping between symmetry-twist sectors}
\label{sec:KTsectormap}

Let us consider how the symmetry-twist sectors are mapped under the  Kennedy-Tasaki transformation \eqref{eq:ktdef}. We again start with the general state $\ket{\psi}$. Assume $\ket{\psi}$ is in the sector labeled by $(u_\sigma, u_\tau, t_\sigma, t_\tau)$, i.e. 
\begin{eqnarray}
\begin{split}
    &\psi_{\{\ssi_i +1, \st_{i-\frac{1}{2}}\}}= (-1)^{u_\sigma} \psi_{\{\ssi_i, \st_{i-\frac{1}{2}}\}}, \hspace{1cm} \psi_{\{\ssi_i , \st_{i-\frac{1}{2}}+1\}}= (-1)^{u_\tau} \psi_{\{\ssi_i, \st_{i-\frac{1}{2}}\}}\\
    & \ssi_{i+L} = \ssi_i + t_\sigma, \hspace{1cm} \st_{i-\frac{1}{2}+L} = \st_{i-\frac{1}{2}}+t_\tau.
\end{split}
\end{eqnarray}
Let us determine the symmetry-twist sectors of the state $\KT\ket{\psi}$ under the Kennedy-Tasaki transformation. To see this, we compute 
\begin{eqnarray}
\begin{split}
    \KT\ket{\psi} = \sum_{\{\sprimesi_{i} , \sprimet_{i-\frac{1}{2}}\}} {\psi'}_{\{\sprimesi_{i} , \sprimet_{i-\frac{1}{2}}\}} \ket{\{\sprimesi_{i} , \sprimet_{i-\frac{1}{2}}\}}
\end{split}
\end{eqnarray}
where 
\begin{eqnarray}
\begin{split}
 {\psi'}_{\{\sprimesi_{i} , \sprimet_{i-\frac{1}{2}}\}}=&\frac{1}{2^{L+1}} \sum_{\{\ssi_{i}, \st_{i-\frac{1}{2}}\}} {\psi}_{\{\ssi_{i} , \st_{i-\frac{1}{2}}\}} (-1)^{\sum_{j=1}^L (\ssi_j + \sprimesi_j)(\st_{j-\frac{1}{2}} + \st_{j+\frac{1}{2}} + \sprimet_{j-\frac{1}{2}}+ \sprimet_{j+\frac{1}{2}}) + (\st_{\frac{1}{2}}+ \sprimet_{\frac{1}{2}})(t_\sigma + t'_{\sigma})}.
\end{split}
\end{eqnarray}
To see $u'_\sigma$ and $u'_\tau$, we compute ${\psi'}_{\{\sprimesi_{i}+1 , \sprimet_{i-\frac{1}{2}}\}}$ and ${\psi'}_{\{\sprimesi_{i} , \sprimet_{i-\frac{1}{2}}+1\}}$ respectively. For ${\psi'}_{\{\sprimesi_{i}+1 , \sprimet_{i-\frac{1}{2}}\}}$, shifting $\sprimesi_i$ by one amounts to multiplying the wavefunction by $(-1)^{t_{\tau}+ t_{\tau}'}$, hence we arrive at 
\begin{eqnarray}
u_\sigma' = t_{\tau}+ t_{\tau}'.
\end{eqnarray}
For ${\psi'}_{\{\sprimesi_{i} , \sprimet_{i-\frac{1}{2}}+1\}}$, shifting $\sprimet_{i-\frac{1}{2}}$ by one amounts to multiplying the wavefunction by $(-1)^{t_\sigma+ t_\sigma'}$, hence we arrive at
\begin{eqnarray}
u_\tau'= t_\sigma+ t_\sigma'.
\end{eqnarray}
On the other hand, shifting $\sprimesi_i$ by one can be undone by shifting $\ssi_i$ by one, because they always come in the combination $\sprimesi_i + \ssi_i$. Hence we also have 
\begin{eqnarray}
u_\sigma'=u_\sigma, \hspace{1cm} u_\tau'=u_\tau.
\end{eqnarray}
Thus the mapping between the symmetry and twist sectors is
\begin{eqnarray}\label{eq:KTsectormaps}
(u_\sigma', u_\tau', t_\sigma', t_\tau')= (u_\sigma, u_\tau, u_\tau+ t_\sigma,   u_\sigma+t_\tau).
\end{eqnarray}
This map is represented in the Table \ref{tab:KTsectors}. 
These mappings among symmetry/twist sectors are exactly
of the same form as \eqref{eq:ux} and~\eqref{eq:tx} for the original Kennedy-Tasaki transformation for $S=1$ chain.

\begin{table}[h]
    \centering
    \renewcommand{\arraystretch}{1.5}
    \begin{tabular}{|c|c|c|c|c|}
    \hline
        $(u_\sigma, u_\tau; t_\sigma, t_\tau)$ & $(0,0)$ & $(1,0)$ & $(0,1)$ & $(1,1)$ \\
        \hline
        $(0,0)$ & \tikzmark{a1} & \tikzmark{a2} & \tikzmark{a3} & \tikzmark{a4} \\
        \hline
        $(1,0)$ & \tikzmark{b1} & \tikzmark{b2} & \tikzmark{b3} & \tikzmark{b4}\\
        \hline
        $(0,1)$ & \tikzmark{c1} & \tikzmark{c2} & \tikzmark{c3} & \tikzmark{c4}\\
        \hline
        $(1,1)$ & \tikzmark{d1} & \tikzmark{d2} & \tikzmark{d3} & \tikzmark{d4}\\
        \hline
    \end{tabular}
    \caption{Mapping between sectors under the Kennedy-Tasaki, i.e. $STS$, transformation.  The rows are labeled by $(u_\sigma,u_\tau)$, and the columns are labeled by $(t_\sigma,t_\tau)$.  The cells without an arrow map to themselves. The two cells connected by an arrow are mapped to each other. }
    \begin{tikzpicture}[overlay, remember picture, shorten >=.5pt, shorten <=.5pt, transform canvas={yshift=.25\baselineskip}]
    \draw [<->] ([yshift=-2pt]{pic cs:b1}) to ([yshift=-2pt]{pic cs:b3});
    \draw [<->] ([yshift=2pt]{pic cs:b2}) -- ([yshift=2pt]{pic cs:b4});
    \draw [<->] ([yshift=0pt]{pic cs:c1})  to ([yshift=0pt]{pic cs:c2});
    \draw [<->] ([yshift=0pt]{pic cs:c3})  to ([yshift=0pt]{pic cs:c4});
    \draw [<->] ([yshift=-2pt]{pic cs:d1})  to ([yshift=-2pt]{pic cs:d4});
    \draw [<->] ([yshift=2pt]{pic cs:d2})  to ([yshift=2pt]{pic cs:d3});
  \end{tikzpicture}
    \label{tab:KTsectors}
\end{table}

\subsection{Non-invertible fusion rules}
\label{sec:KTfusionrule}

We proceed to discuss the fusion rule involving $\KT, U_\sigma$ and $U_\tau$. We first consider the fusion rule $\KT\times U_\sigma$. This has already been discussed in the previous subsection. We first note that by definition of \eqref{eq:Z2Z2op},
\begin{eqnarray}
\CN_{\text{KT}} \ket{\{\ssi_{i}+1, \st_{i-\frac{1}{2}}\}} = \CN_{\text{KT}} U_{\sigma} \ket{\{\ssi_{i}, \st_{i-\frac{1}{2}}\}}.
\end{eqnarray}
On the other hand, by the definition of the $\KT$ \eqref{eq:ktdef}, shifting $\ssi_i$ by 1 amounts to multiplying 
\begin{eqnarray}
(-1)^{\sum_{j=1}^L \st_{j-\frac{1}{2}} + \st_{j+\frac{1}{2}} + \sprimet_{j-\frac{1}{2}}+ \sprimet_{j+\frac{1}{2}}} = (-1)^{t_{\tau} + {t}'_{\tau}}.
\end{eqnarray}
Hence
\begin{eqnarray}
\CN_{\text{KT}} U_{\sigma} \ket{\{\ssi_{i}, \st_{i-\frac{1}{2}}\}}= (-1)^{t_{\tau} + {t}'_{\tau}}\KT \ket{\{\ssi_{i}, \st_{i-\frac{1}{2}}\}}.
\end{eqnarray}
This justifies the fusion rule 
\begin{eqnarray}
\KT \times U_{\sigma} = (-1)^{t_\tau + t_\tau'} \KT.
\end{eqnarray}
By a similar calculation, we also find that 
\begin{eqnarray}
\KT \times U_{\tau} = (-1)^{t_\sigma + t_\sigma'} \KT.
\end{eqnarray}

We further compute the fusion rule $\KT\times \KT$. By definition, we have
\begin{equation}
\begin{split}
    &\KT\times \KT \ket{\{\ssi_{i}, \st_{i-\frac{1}{2}}\}}\\ &= \frac{1}{4^{L-1}} \sum_{\{\sprimesi_{i}, \sprimet_{i-\frac{1}{2}}, \sprimeprimesi_{i}, \sprimeprimet_{i-\frac{1}{2}}\}} (-1)^{\sum_{j=1}^L (\ssi_j + \sprimesi_j)(\st_{j-\frac{1}{2}} + \st_{j+\frac{1}{2}} + \sprimet_{j-\frac{1}{2}}+ \sprimet_{j+\frac{1}{2}}) + (\st_{\frac{1}{2}}+ \sprimet_{\frac{1}{2}})(t_\sigma + t'_{\sigma})}\\
    &\hspace{1cm}\times(-1)^{\sum_{j=1}^L (\sprimeprimesi_j + \sprimesi_j)(\sprimeprimet_{j-\frac{1}{2}} + \sprimeprimet_{j+\frac{1}{2}} + \sprimet_{j-\frac{1}{2}}+ \sprimet_{j+\frac{1}{2}}) + (\sprimeprimet_{\frac{1}{2}}+ \sprimet_{\frac{1}{2}})(t_\sigma'' + t'_{\sigma})}
    \ket{\{\sprimeprimesi_i, \sprimeprimet_{i-\frac{1}{2}} \}}.
\end{split}
\end{equation}
We first sum over $\sprimesi_j$ which enforces $\st_{j-\frac{1}{2}}+ \st_{j+\frac{1}{2}} + \sprimeprimet_{j-\frac{1}{2}}+ \sprimeprimet_{j+\frac{1}{2}}=0$ mod 2. Solving this enforces $\st_{j-\frac{1}{2}}+ \sprimeprimet_{j-\frac{1}{2}}$ to be a constant $c_\tau$. In particular, this constraints $t_\tau= t_\tau''$. We further sum over $\sprimet_{j-\frac{1}{2}}$ which enforces $\ssi_{j}+ \sprimeprimesi_{j}$ to be a constant $c_\sigma$. In particular, this constraints $t_\sigma=t_\sigma''$. Finally summing over $c_\tau, c_\sigma\in \Z_2$, we find
\begin{eqnarray}
\begin{split}
    &\KT\times \KT \ket{\{\ssi_{i}, \st_{i-\frac{1}{2}}\}}= 4\left( 1+ (-1)^{t_\sigma+ t_\sigma'} U_\tau\right)  \left( 1+ (-1)^{t_\tau+ t_\tau'} U_\sigma\right) 
    \ket{\{\ssi_i, \st_{i-\frac{1}{2}} \}},
\end{split}
\end{eqnarray}
 which implies the fusion rule
\begin{eqnarray}\label{eq:KTKTfusionrule}
\begin{split}
    \KT\times \KT =4\left( 1+ (-1)^{t_\sigma+ t_\sigma'} U_\tau\right)  \left( 1+ (-1)^{t_\tau+ t_\tau'} U_\sigma\right).
\end{split}
\end{eqnarray}
Same comments in Section \ref{sec:KWfusion} also apply here. The non-trivial signs mean that the presence of additional $\Z_2\times \Z_2$ operators terminating on $\KT$'s modifies the fusion rule. Usually when we refer to the fusion rule, we assume $\Z_2\times \Z_2$ operators are all turned off nearby $\KT$'s, hence $t_{\sigma,\tau}=t_{\sigma,\tau}'=0$. After this simplification, the fusion rule \eqref{eq:KTKTfusionrule} is almost identical to the standard fusion rule of the $\Z_2\times \Z_2$ Tambara-Yamagami (TY) fusion category. The only difference is the phase $4$ on the right hand side. It means that our $\KT$ is stacking of the duality defect in the $\Z_2\times \Z_2$ TY category by a $(0+1)$d $\Z_2$ TQFT, i.e. a two level quantum mechanics. Hence our $\KT$ is a non-simple operator, but a sum of two simple operators. Such a simple operator turns out to implement the $TST$ (rather than $STS$) transformation. \footnote{Note that the line operator implementing $TST$ is $U_{\text{DW}}\CN U_{\text{DW}}$ which has quantum dimension 2. On the other hand the defect implementing $STS$ is given by $\KT= \CN^{\dagger} U_{\text{DW}} \CN$ which has quantum dimension 4. Indeed, their quantum dimensions differ by 2, suggesting that the later is stacking a $\Z_2$ TQFT onto the former. }

The most important information from the fusion rule \eqref{eq:KTKTfusionrule} is that $\KT$ is a non-invertible operator satisfying the non-invertible fusion rule, and hence the Kennedy-Tasaki transformation associated with $STS$ is a non-unitary transformation. In particular, $\KT$ annihilates any state that is odd under any one of the $\Z_2$'s, i.e. has eigenvalue $-1$ under $U_\tau$ or $U_\sigma$, under the periodic boundary condition ($t_{\sigma,\tau}=t_{\sigma,\tau}'=0$). 
As we discussed in Section~\ref{sec:KTspin1ring} for the $S=1$ chain, we may also interpret $\KT$ as an unitary operator acting on
an \emph{extended} Hilbert space.

Despite the disadvantage that the operator $\KT$ being non-simple, we still prefer $STS$ over $TST$ in our definition, for which the reason will become clear once we formulate the Kennedy-Tasaki transformation for spin-$\frac{1}{2}$ system on an open chain.

\subsection{Kennedy-Tasaki and $\Z_2\times \Z_2$ twisted gauging  }

We have derived the map between symmetry and twist sectors under the Kennedy-Tasaki transformation implementing $STS$, on the lattice. In this subsection, we derive the sector mapping using the partition function and the definition of gauging. We start with the partition function of theory $\CX$ with a non-anomalous $\Z_2\times \Z_2$ global symmetry whose background fields are $A_1, A_2$, i.e. $Z_{\CX}[A_1, A_2]$. The $STS$ transformation acts on the partition function as
\begin{eqnarray}
\begin{split}
    Z_{STS\CX}[A_1', A_2']& = \frac{1}{|H^0(X_2, \Z_2)|^4} \sum_{a_1, a_2, \widehat{a}_1, \widehat{a}_2} Z_{\CX}[a_1, a_2] (-1)^{\int_{X_2} a_1 \widehat{a}_2 + a_2 \widehat{a}_1 + \widehat{a}_1 \widehat{a}_2 + A_1' \widehat{a}_2 + A_2' \widehat{a}_1 }\\
    &= \frac{|H^1(X_2, \Z_2)|}{|H^0(X_2, \Z_2)|^4}\sum_{a_1, a_2} Z_{\CX}[a_1, a_2] (-1)^{\int_{X_2} a_1 {a}_2 +  A_1' {a}_2 + A_2' {a}_1 + A_1' A_2' }
\end{split}
\end{eqnarray}
where $a_1$ and $a_2$ ($\widehat{a}_1$ and $\widehat{a}_2$) are dynamical gauge field of $\Z_2\times \Z_2$ in first (second) $S$ transformation. $H^0(X_2,\Z_2)$ and $H^1(X_2,\Z_2)$ are first and second cohomology on manifold $X_2$ with $\Z_2$ coefficient.

To define the symmetry and twist sectors, we formulate the theory on the torus, and summing over gauge fields reduces to summing over holonomies around the two non-contractible cycles.
\begin{eqnarray}
\begin{split}
    Z_{STS\CX}[W'^{t}_{1}, W'^{t}_2, W'^{x}_{1}, W'^{x}_{2}]=& \frac{1}{4} \sum_{w_1^t, w_2^t, w_1^x, w_2^x} Z_{\CX}[w_1^t, w_2^t, w_1^x, w_2^x] (-1)^{w_1^t w_2^x+ w_1^x w_2^t }\\
    &(-1)^{w_1^t W'^{x}_2 + w_1^x W'^{t}_2 + w_2^t W'^{x}_1 + w_2^x W'^{t}_1+ W'^{x}_1 W'^{t}_2 + W'^{t}_{1} W'^{x}_2}.
\end{split}
\end{eqnarray}
The partition function in terms of the holonomies and in terms of the symmetry-twist sectors are related via 
\begin{eqnarray}
Z_{\CX}^{(u_\sigma, u_\tau, t_\sigma, t_\tau)} = \frac{1}{4} \sum_{w_1^t, w_2^t} Z_{\CX}[w_1^t, w_2^t, t_\sigma, t_\tau] (-1)^{u_\sigma w_1^t + u_\tau w_2^t}
\end{eqnarray}
and the inverse relation is 
\begin{eqnarray}
Z_{\CX}[w_1^t, w_2^t, w_1^x, w_2^x]= \sum_{u_\sigma, u_\tau} Z_{\CX}^{(u_\sigma, u_\tau, w_1^x, w_2^x)} (-1)^{u_\sigma w_1^t + u_\tau w_2^t}.
\end{eqnarray}
Combining the above relations, we find the desired relation 
\begin{eqnarray}
Z_{STS\CX}^{(u_\sigma', u_\tau', t_\sigma', t_\tau')} = Z_{\CX}^{(u_\sigma', u_\tau', u_\tau'+ t_\sigma', u_\sigma'+ t_\tau')} := Z_{\CX}^{(u_\sigma, u_\tau, t_\sigma, t_\tau)}.
\end{eqnarray}
This means that $(u_\sigma', u_\tau', u_\tau'+ t_\sigma', u_\sigma'+ t_\tau')= (u_\sigma, u_\tau, t_\sigma, t_\tau)$ which is equivalent to \eqref{eq:KTsectormaps} as well as \eqref{eq:ux} and \eqref{eq:tx} for the spin-1 system.

The fusion rule can also be reproduced from the partition function approach. The fusion rules for the defect implementing $TST$ have been worked out in \cite{Kaidi:2021xfk}. The similar calculation for $STS$ can be worked out as well. We will not repeat the exercise here.

\section{Kennedy-Tasaki transformation on an interval with two spin-$\frac{1}{2}$'s per unit cell: A unitary transformation}
\label{sec:KTopenbdy}

We proceed to discuss the Kennedy-Tasaki transformation for spin-$\frac{1}{2}$ system on an open chain.
While similar transformations on an open chain were discussed earlier~\cite{PhysRevB.46.3486,takada1992nonlocal},
our construction clarifies its connection to various modern concepts related to SPT phases.
Similarly to the Kramers-Wannier transformation, we will find that although the operator $\KT$ implementing $STS$ is non-unitary and satisfies non-invertible fusion rule, the $\KT$ under the free open boundary condition is a unitary operator.

Suppose the open chain contains sites at coordinate $i$ and links at coordinate $i-\frac{1}{2}$, with  $i=1, ..., L$. We begin by modifying \eqref{eq:ktdef} such that only the terms that are fully supported on the chain will be kept in the exponent, i.e. free boundary condition.   Concretely, 
\begin{equation}\label{eq:KTopenchain}
\begin{split}
    \CN_{\text{KT}}^{\text{open}} \ket{\{\ssi_i, \st_{i-\frac{1}{2}} \}} = \frac{1}{2^{L}} \sum_{\{\sprimesi_{i}, \sprimet_{i-\frac{1}{2}}\}}(-1)^{\sum_{j=1}^L (\ssi_j + \sprimesi_j)(\st_{j-\frac{1}{2}} +  \sprimet_{j-\frac{1}{2}})}(-1)^{\sum_{j=1}^{L-1} (\ssi_j + \sprimesi_j)( \st_{j+\frac{1}{2}} + \sprimet_{j+\frac{1}{2}})} \ket{\{\sprimesi_i, \sprimet_{i-\frac{1}{2}} \}}.
\end{split}
\end{equation}
We use the superscript to distinguish $\KT^{\text{open}}$ defined on an interval from the $\KT$ defined on a ring. 
To check that it is a unitary transformation, we simply consider the overlap, 
\begin{eqnarray}
\begin{split}
    \braket{\{\ssi_i, \st_{i-\frac{1}{2}} \} | \KT^{\text{open} \dagger} \KT^{\text{open}}  | \{\sprimesi_i, \sprimet_{i-\frac{1}{2}} \}}= 
\prod_{j=1}^L\delta_{\ssi_j, \sprimesi_j} \delta_{\st_{j-\frac{1}{2}}, \sprimet_{j-\frac{1}{2}}}.
\end{split}
\end{eqnarray}
Hence $\KT^{\text{open}} $ is unitary and invertible, whose inverse is $\KT^{\text{open}\dagger}$.

Let us examine the Kennedy-Tasaki transformation of spin operators on the open chain.
We can immediately see that the $x$-component of the spin operators $\sigma^x_j, \tau^x_{j-\frac{1}{2}}$
are invariant under the Kennedy-Tasaki transformation:
\begin{eqnarray}
    \CN_{\text{KT}}^{\text{open}} \sigma^x_j {\CN_{\text{KT}}^{\text{open}}}^\dagger &=& \sigma^x_j ,
    \label{eq:sigmax_openKT}
    \\
   \CN_{\text{KT}}^{\text{open}} \tau^x_{j-\frac{1}{2}} {\CN_{\text{KT}}^{\text{open}}}^\dagger &=& \tau^x_{j-\frac{1}{2}} ,
\end{eqnarray}
since they are mapped to diagonal operators in $\tau^z, \sigma^z$-basis
by the Kramers-Wannier transformation $\CN$. Since these diagonal operators commute with $U_{\text{DW}}$, they are mapped back to the
original operators by $\CN^\dagger$.
On the other hand, using the Kramers-Wannier transformation~\eqref{eq:tauz_sigmax} of the spin operators on the open chain,
and the transformation~\cite{scaffidi2017gapless} by $U_{\text{DW}}$
\begin{eqnarray}
    U_{\text{DW}}  \tau^x_{j-\frac{1}{2}} U_{\text{DW}}^\dagger &=& 
\begin{cases}
    \sigma^z_{j-1} \tau^x_{j-\frac{1}{2}} \sigma^z_j & ( j= 2,3,\ldots,L) ,
    \\
    \tau^x_{\frac{1}{2}} \sigma^z_1 & (j=1) ,
\end{cases}
\\
    U_{\text{DW}}  \sigma^x_j U_{\text{DW}}^\dagger &=& 
\begin{cases}
    \tau^z_{j-\frac{1}{2}} \sigma^x_{j} \tau^z_{j+\frac{1}{2}} & ( j= 1,2,\ldots,L-1) ,
    \\
    \tau^z_{L-\frac{1}{2}} \sigma^x_L & (j=L) ,
\end{cases}
\end{eqnarray}
and that $\sigma^z,\tau^z$ are unchanged by $U_{\text{DW}}$ as mentioned above, we find
\begin{eqnarray}
    \CN_{\text{KT}}^{\text{open}} \sigma^z_j {\CN_{\text{KT}}^{\text{open}}}^\dagger &=& \left( \prod_{k=1}^j \tau^x_{k-\frac{1}{2}} \right) \sigma^z_j ,
  \label{eq:sigmaz_openKT}
    \\
   \CN_{\text{KT}}^{\text{open}} \tau^z_{j-\frac{1}{2}} {\CN_{\text{KT}}^{\text{open}}}^\dagger &=& \tau^z_{j-\frac{1}{2}} \left( \prod_{k=j}^L \sigma^x_k \right) .
 \label{eq:tauz_openKT}
\end{eqnarray}

As a consequence of Eq.~\eqref{eq:sigmax_openKT}, the symmetry generators~\eqref{eq:Z2Z2op} are also invariant. 
This feature that the symmetry is preserved under the Kennedy-Tasaki transformation, as it is the case in the original Kennedy-Tasaki transformation for spin-1 systems,
is particularly convenient, and holds for $STS$ but not for $TST$.
For $TST$ on an open chain, the symmetry operator will be mapped to a local operator, which is not the case for the original Kennedy-Tasaki transformation.
This is the main reason we prefer $STS$ over $TST$.

\section{Gapped SPT in spin-$\frac{1}{2}$ system from Kennedy-Tasaki transformation}
\label{sec:gappedSPT}

The Kennedy-Tasaki transformation was designed to map a $\Z_2\times \Z_2$ symmetry spontaneously broken (SSB) phase to a $\Z_2\times \Z_2$ symmetry protected topological (SPT) phase.  It is straightforward to check at the level of partition function that  $STS$ transformation relates the two, as shown in Section \ref{sec:fieldtheory}. We will review how the SPT phase can be generated from the Kennedy-Tasaki transformation for the spin-$\frac{1}{2}$ system.

The Hamiltonian for the $\Z_2\times \Z_2$ SSB phase is 
\begin{eqnarray}\label{eq:SSBphase}
H_{\text{SSB}}= -\sum_{i=1}^L \left(\sigma^z_{i-1}\sigma^z_{i}+ \tau^z_{i-\frac{1}{2}} \tau^z_{i+\frac{1}{2}}\right)
\end{eqnarray}
where the degrees of freedom charged under two $\Z_2$'s are decoupled.
Now we apply the Kennedy-Tasaki transformation.

On an open chain of sites $1,2,\ldots,L$, the Hamiltonian reads
\begin{eqnarray}
H_{\text{SSB}}^{\text{open}} = -\sum_{i=2}^L \sigma^z_{i-1}\sigma^z_{i} - \sum_{i=1}^{L-1} \tau^z_{i-\frac{1}{2}} \tau^z_{i+\frac{1}{2}} .
\end{eqnarray}
Using Eqs.~\eqref{eq:sigmaz_openKT} and~\eqref{eq:tauz_openKT}, we find
\begin{eqnarray}
 H_{\text{SPT}}^{\text{open}} & = & \KT^{\text{open}} H_{\text{SSB}}^{\text{open}} {\KT^{\text{open}}}^\dagger
 \nonumber \\
 &=& -\sum_{j=2}^L \sigma^z_{j-1} \tau^x_{j-\frac{1}{2}} \sigma^z_{j} - \sum_{j=1}^{L-1} \tau^{z}_{j-\frac{1}{2}} \sigma^x_{j} \tau^z_{j+\frac{1}{2}} .
\label{eq:cluster_open}
\end{eqnarray}
On the ring, the Kennedy-Tasaki dual of the SSB Hamiltonian~\eqref{eq:SSBphase} is given by
\begin{eqnarray}\label{eq:gappedSPT}
 H_{\text{SPT}}= -\sum_{j=1}^L \left( \sigma^z_{j-1} \tau^x_{j-\frac{1}{2}} \sigma^z_{j} + \tau^{z}_{j-\frac{1}{2}} \sigma^x_{j} \tau^z_{j+\frac{1}{2}} \right),
\end{eqnarray}
with the boundary conditions as discussed in Sec.~\ref{sec:KTclosedbdy}.

The resulting Hamiltonian is precisely the cluster model describing the $\Z_2\times \Z_2$ gapped SPT~\cite{chen2014symmetry}.
For the sake of completeness, here we review how the edge states arise in the cluster model 
defined on an open chain~\eqref{eq:cluster_open} \cite{PhysRevB.91.155150}.
The Hamiltonian~\eqref{eq:cluster_open} is a sum of commuting projectors.
Thus, within the ground-state subspace, all the projectors have eigenvalue one:
\begin{eqnarray}
 \sigma^z_{j-1} \tau^x_{j-\frac{1}{2}} \sigma^z_j \sim 1 && (j=2,3, \ldots,L) ,
\\
 \tau^z_{j-\frac{1}{2}} \sigma^x_j \tau^z_{j+\frac{1}{2}} \sim 1 && (j=1,2,\ldots,L-1) .
\end{eqnarray}
Using these relations, the symmetry generators~\eqref{eq:Z2Z2op} of the $\Z_2 \times \Z_2$ symmetry can be
rewritten (within the ground-state subspace) as $U_{\sigma,\tau} \sim U^L_{\sigma,\tau} \otimes U^R_{\sigma,\tau}$,
where
\begin{eqnarray}
 U^L_\sigma & = &  \tau^z_{\frac{1}{2}},
 \\
 U^R_\sigma &=& \tau^z_{L-\frac{1}{2}}\sigma^x_L ,
 \\
 U^L_\tau & = & \tau^x_{\frac{1}{2}} \sigma^z_1 ,
 \\
 U^R_\tau &=& \sigma^z_L .
\end{eqnarray}
Thus, within the ground-space subspace, symmetry generators effectively act only at the localized
regions near the ends of the chain.
Since the localized symmetry generators at each end anticommute ($U^a_\sigma U^a_\tau = - U^a_\tau U^a_\sigma$
for $a=L,R$), there must be a localized edge state producing two-fold degeneracy, at each end.

\section{Equivalence between Kennedy-Tasaki transformations in spin-1 and spin-$\frac{1}{2}$ systems}
\label{sec:projtospin1}

Finally, we discuss the relation between our Kennedy-Tasaki transformation $\KT$ in \eqref{eq:KTopenchain} for spin-$\frac{1}{2}$ systems  and the original Kennedy-Tasaki transformation for spin-1 systems, both on a ring and on an interval. 
\begin{enumerate}
    \item On a ring, the Kennedy-Tasaki transformation for spin-$\frac{1}{2}$ system $\KT$ and that for spin-1 system as defined in \eqref{eq:Sx_KT1} are equivalent.
    \item On an interval, the Kennedy-Tasaki transformation for spin-$\frac{1}{2}$ system $\KT^{\text{open}}$ and  the original non-local unitary operator $U_{\text{KT}}$ for spin-1 systems are almost equivalent, up to a symmetry sector dependent sign. This sign is potentially due to the choice of boundary conditions. 
\end{enumerate}

\subsection{Relating the Hilbert space of spin-1 and two spin-$\frac{1}{2}$'s}

The Hilbert space for each spin-1 is three dimensional, whose basis states are denoted as $\ket{+}, \ket{0}$ and $\ket{-}$. The Hilbert space for two spin-$\frac{1}{2}$'s is four dimensional, whose basis stats are denoted as $\ket{\uparrow\uparrow}, \ket{\uparrow\downarrow}, \ket{\downarrow\uparrow}$ and $\ket{\downarrow\downarrow}$. To make a connection with the two basis, we start with the spin-1 basis and bring in another spin-0 state to make a four dimensional Hilbert space. The basis states are mapped as follows,
\begin{eqnarray}
\text{spin-}1: \hspace{1cm} 
\begin{cases}
\ket{+} = \ket{\uparrow \uparrow}\\
\ket{0} = \frac{1}{\sqrt{2}} (\ket{\uparrow \downarrow}+ \ket{\downarrow \uparrow})\\
\ket{-} = \ket{\downarrow \downarrow}\\
\end{cases};\hspace{1.5cm}
\text{spin-}0: \hspace{1cm} 
\frac{1}{\sqrt{2}} (\ket{\uparrow \downarrow}- \ket{\downarrow \uparrow})
\end{eqnarray}
where the spin-0 state is the additional state not belonging to the original Hilbert space. In terms of Pauli operators, we have 
\begin{eqnarray}\label{eq:spin11/2}
S^{x}_i= \frac{1}{2} (\sigma_i^x + \tau^x_{i-\frac{1}{2}}),\hspace{1cm} S^{z}_i= \frac{1}{2} (\sigma_i^z + \tau^z_{i-\frac{1}{2}}).
\end{eqnarray}
The symmetry generators of $\Z_2\times \Z_2$  are
$R^{x,z}$ defined in Eq.~\eqref{eq:defR},
which correspond to $\pi$ rotations around the $x$ and $z$ axises respectively. In terms of spin-$\frac{1}{2}$ variables, the two symmetry generators are given by 
\begin{eqnarray}\label{eq:Z2Z21}
R^x=e^{\frac{i \pi}{2} \sum_{j=1}^{L} (\sigma_j^x + \tau^x_{j-\frac{1}{2}})} = (-1)^L \prod_{j=1}^L \sigma_j^x \tau_{j-\frac{1}{2}}^x, \hspace{1cm} R^z=e^{\frac{i \pi}{2} \sum_{j=1}^{L} (\sigma_j^z + \tau^z_{j-\frac{1}{2}})} = (-1)^L \prod_{j=1}^L \sigma_j^z \tau_{j-\frac{1}{2}}^z.
\end{eqnarray}

We would like to further identify $R^x=U_\sigma, R^z=U_\tau$ where $U_{\sigma,\tau}$ are defined in \eqref{eq:Z2Z2op}. To achieve this, we need to perform a basis rotation such that the Pauli operators are mapped as follows, 
\begin{eqnarray}\label{eq:basisrotation}
\begin{pmatrix}
\sigma_i^x\\
\sigma_i^z\\
\tau_{i-\frac{1}{2}}^x\\
\tau_{i-\frac{1}{2}}^z
\end{pmatrix} \to 
\begin{pmatrix}
-\sigma^x_i \tau^z_{i-\frac{1}{2}}\\
\sigma_i^z\\
\tau^z_{i-\frac{1}{2}}\\
-\sigma_i^z \tau^x_{i-\frac{1}{2}}.
\end{pmatrix}
\end{eqnarray}
In terms of the rotated Pauli operators, the above $\Z_2\times \Z_2$ symmetry generators are indeed standard ones, $\prod_{j=1}^L \sigma^x_j$ and $\prod_{j=1}^{L} \tau^x_{j-\frac{1}{2}}$.

\subsection{Equivalence on a ring}

We proceed to show that when defined on a ring,  the Kennedy-Tasaki transformation for spin-$\frac{1}{2}$ systems, i.e. $\KT$, is equivalent to that for spin-1 systems, which was defined in Section \ref{sec:KTspin1ring}.

Recall that for spin-1 systems on a ring, we defined the Kennedy-Tasaki transformation via specifying how the spin operators transform, as shown in \eqref{eq:Sx_KT1}. To compare it with $\KT$ defined in \eqref{eq:ktdef}, we first derive how the spin-$\frac{1}{2}$ Pauli operators transform on a ring. This can be achieved by showing the following identities hold when acting on arbitrary basis states $\ket{\{s^\sigma_{i}, s^\tau_{i-\frac{1}{2}}\}}$, 
\begin{eqnarray}\label{eq:KTring}
\begin{split}
    \KT \sigma^z_{i} &= (-1)^{t_\sigma+t'_\sigma} \prod_{j=1}^{i} \tau'^x_{j-\frac{1}{2}} \sigma'^z_{i} \KT,\\
    \KT \tau^z_{i-\frac{1}{2}} &= (-1)^{t_\tau+t'_\tau} \prod_{j=i}^{L} \sigma'^x_{j} \tau'^z_{i-\frac{1}{2}} \KT,\\
    \KT \sigma^x_{i} &= \sigma'^x_{i} \KT,\\
    \KT \tau^x_{i} &= \tau'^x_{i} \KT.\\
\end{split}
\end{eqnarray}
In brief, we have $\sigma^z_{i}= (-1)^{t_\sigma+t'_\sigma} \prod_{j=1}^{i} \tau'^x_{j-\frac{1}{2}} \sigma'^z_{i} , \tau^z_{i-\frac{1}{2}}= (-1)^{t_\tau+t'_\tau} \prod_{j=i}^{L} \sigma'^x_{j} \tau'^z_{i-\frac{1}{2}}, \sigma^x_{i}=\sigma'^x_{i}$ and $\tau^x_{i} =\tau'^x_{i}$. 
Let us check how \eqref{eq:spin11/2} and \eqref{eq:basisrotation} together with \eqref{eq:Sx_KT1} reproduces \eqref{eq:KTring}. We first note that combination of \eqref{eq:spin11/2} and \eqref{eq:basisrotation} gives $S^x_j= \frac{1}{2}\tau^z_{j-\frac{1}{2}}(1-\sigma^x_{j})$ and $S^z_j= \frac{1}{2}\sigma^z_{j}(1-\tau^x_{j-\frac{1}{2}})$. We then start with \eqref{eq:Sx_KT1}, 
\begin{eqnarray}
\begin{split}
     \frac{1}{2}\tau^z_{j-\frac{1}{2}}(1-\sigma^x_{j})\stackrel{}{=} S^x_{j} \stackrel{\eqref{eq:Sx_KT1}}{=} e^{i\pi \sum_{k=1}^{j-1} {S'}_k^x} {S'}_j^x &= e^{i\pi \sum_{k=1}^{j-1} \frac{1}{2}{\tau'}^z_{k-\frac{1}{2}}(1-\sigma'^x_k)} \frac{1}{2}\tau'^z_{j-\frac{1}{2}}(1-\sigma'^x_j)\\
    &=\left(\prod_{k=1}^{j-1} {\sigma'}_k^x\right) \frac{1}{2}\tau'^z_{j-\frac{1}{2}}(1-\sigma'^x_j)\\
    &= U'_\sigma \left(\prod_{k=j}^{L} {\sigma'}_k^x\right) \frac{1}{2}\tau'^z_{j-\frac{1}{2}}(1-\sigma'^x_j)\\
    &= (-1)^{u'_\sigma} \left(\prod_{k=j}^{L} {\sigma'}_k^x\right) \frac{1}{2}\tau'^z_{j-\frac{1}{2}}(1-\sigma'^x_j).
\end{split}
\end{eqnarray}
Comparing the first and the last expression, and using the relation $t_\tau+t'_\tau=u'_\sigma$, this is nothing but the second identity in \eqref{eq:KTring}. Similarly, we can also get
\begin{eqnarray}
\begin{split}
     \frac{1}{2}\sigma^z_{j}(1-\tau^x_{j-\frac{1}{2}})\stackrel{}{=} S^z_{j} \stackrel{\eqref{eq:Sx_KT1}}{=} e^{i\pi \sum_{k=j+1}^{L} {S'}_k^z} {S'}_j^z &= e^{i\pi \sum_{k=j+1}^{L} \frac{1}{2}\sigma'^z_{k}(1-\tau'^x_{k-\frac{1}{2}})} \frac{1}{2}\sigma'^z_{j}(1-\tau'^x_{j-\frac{1}{2}})\\
    &=\left(\prod_{k=j+1}^{L} {\tau'}_{k-\frac{1}{2}}^x\right) \frac{1}{2}\sigma'^z_{j}(1-\tau'^x_{j-\frac{1}{2}})\\
    &= U'_\tau \left(\prod_{k=1}^{j-1} {\tau'}_{k-\frac{1}{2}}^x\right) \frac{1}{2}\sigma'^z_{j}(1-\tau'^x_{j-\frac{1}{2}})\\
    &= (-1)^{u'_\tau} \left(\prod_{k=1}^{j-1} {\tau'}_{k-\frac{1}{2}}^x\right) \frac{1}{2}\sigma'^z_{j}(1-\tau'^x_{j-\frac{1}{2}})
\end{split}
\end{eqnarray}
which, upon using $u_\tau'=t_\sigma+t_\sigma'$, reproduces the first equality in \eqref{eq:KTring}. This establishes that upon introducing the fourth spin-0 state, the Kennedy-Tasaki transformation for spin-1 systems on a ring is equivalent to that for spin-$\frac{1}{2}$ systems.

\subsection{Almost equivalence on an interval}

We finally proceed to show that $\KT^{\text{open}}$ and $U_{\text{KT}}$ are almost equivalent up to a sign depending on the symmetry sectors on an interval. Since on a closed chain they are equivalent, this subtle sign may potentially come from the different choice of boundary conditions. Below, we first recast the $\KT^{\text{open}}$ as a unitary operator in terms of Pauli operators, and then compare it with $U_{\text{KT}}$ via the map \eqref{eq:spin11/2} and \eqref{eq:basisrotation}. 

\paragraph{Recasting $\KT^{\text{open}}$ as an explicit unitary operator:}
The  Kennedy-Tasaki transformation \eqref{eq:KTopenchain} was defined via specifying how $\KT$ acts on the Hilbert space. It is useful to write down the operator $\KT$ in terms of Pauli operators explicitly.

As a first step, it is useful to note that $\KT$ is the composition of ${\CN}^\dagger U_{\text{DW}} \CN$, which implements $STS$. Here $\CN^\dagger= \CN^{-1}$ on an open chain implementing Kramers-Wannier transformations.  Note that the two $S$'s in $STS$ act on different Hilbert spaces, the second acts on the original one, and the first acts on the dual one. On a closed chain $S$ is self-conjugate, hence we don't distinguish ${S}^\dagger$ and $S$. However, on an open chain ${S}^\dagger$ and $S$ are different (dispite $S$ is unitary), and correspondingly ${\CN}^\dagger$ and $\CN$ are also different. For simplicity, on the open chain, we still write the operation as $STS$, but should keep in mind that it is implemented by the operator $\CN^\dagger U_{\text{DW}} \CN$. 

We can check $\KT^{\text{open}}= \CN^\dagger U_{\text{DW}} \CN$ explicitly. We first have 
\begin{equation}\label{eq:KTNDWN}
\begin{split}
    \CN\ket{\{\ssi_{j}, \st_{j-\frac{1}{2}} \}} = \frac{1}{2^L} \sum_{\{\hssi_{j-\frac{1}{2}}, \hst_{j} \}} (-1)^{\sum_{j=2}^L (\ssi_{j-1} + \ssi_j) \hssi_{j-\frac{1}{2}} + \ssi_1 \hssi_{\frac{1}{2}} }  (-1)^{\sum_{j=1}^{L-1} (\st_{j-\frac{1}{2}}+ \st_{j+\frac{1}{2}}) \hst_{j} + \st_{L-\frac{1}{2}} \hst_L} \ket{\{ \hssi_{j-\frac{1}{2}}, \hst_{j} \}}.
\end{split}
\end{equation}
Its Hermitian conjugate is 
\begin{equation}
\begin{split}
    \CN^\dagger\ket{\{ \hssi_{j-\frac{1}{2}}, \hst_{j} \}} = \frac{1}{2^L} \sum_{\{\ssi_{j}, \st_{j-\frac{1}{2}} \}} (-1)^{\sum_{j=2}^L (\ssi_{j-1} + \ssi_j) \hssi_{j-\frac{1}{2}} + \ssi_1 \hssi_{\frac{1}{2}} }  (-1)^{\sum_{j=1}^{L-1} (\st_{j-\frac{1}{2}}+ \st_{j+\frac{1}{2}}) \hst_{j} + \st_{L-\frac{1}{2}} \hst_L} \ket{\{\ssi_{j}, \st_{j-\frac{1}{2}} \}}.
\end{split}
\end{equation}
Then we can consider the product, 
\begin{eqnarray}
\begin{split}
    &\CN^\dagger U_{\text{DW}} \CN \ket{\{\ssi_{j}, \st_{j-\frac{1}{2}} \}} \\
    &= \frac{1}{4^L} \sum_{\{\hssi_{j-\frac{1}{2}}, \hst_{j} \}, \{\sprimesi_{j}, \sprimet_{j-\frac{1}{2}} \}} (-1)^{\sum_{j=2}^L (\ssi_{j-1} + \ssi_j) \hssi_{j-\frac{1}{2}} + \ssi_1 \hssi_{\frac{1}{2}} }  (-1)^{\sum_{j=1}^{L-1} (\st_{j-\frac{1}{2}}+ \st_{j+\frac{1}{2}}) \hst_{j} + \st_{L-\frac{1}{2}} \hst_L}\\
    & \hspace{1cm} \times  (-1)^{\sum_{j=1}^{L-1} \hst_{j}(\hssi_{j-\frac{1}{2}} + \hssi_{j+\frac{1}{2}} ) + \hssi_{L-\frac{1}{2}} \hst_{L}  }  (-1)^{\sum_{j=2}^L (\sprimesi_{j-1} + \sprimesi_j) \hssi_{j-\frac{1}{2}} + \sprimesi_1 \hssi_{\frac{1}{2}} }  \\&\hspace{1cm}\times  (-1)^{\sum_{j=1}^{L-1} (\sprimet_{j-\frac{1}{2}}+ \sprimet_{j+\frac{1}{2}}) \hst_{j} + \sprimet_{L-\frac{1}{2}} \hst_L}\ket{\{\sprimesi_{j}, \sprimet_{j-\frac{1}{2}} \}}.
\end{split}
\end{eqnarray}
Summing over $\hst_{j}$ for any $j$ yields $\hssi_{j-\frac{1}{2}}= \st_{j-\frac{1}{2}} + \sprimesi_{j-\frac{1}{2}}$ for any $j$. Substituting the solution into the above equation, we find that 
\begin{eqnarray}
\begin{split}
    &\CN^\dagger U_{\text{DW}} \CN \ket{\{\ssi_{j}, \st_{j-\frac{1}{2}} \}} \\&= \frac{1}{2^{L}} \sum_{\{\sprimesi_{i}, \sprimet_{i-\frac{1}{2}}\}}(-1)^{\sum_{j=1}^L (\ssi_j + \sprimesi_j)(\st_{j-\frac{1}{2}} +  \sprimet_{j-\frac{1}{2}})}(-1)^{\sum_{j=1}^{L-1} (\ssi_j + \sprimesi_j)( \st_{j+\frac{1}{2}} + \sprimet_{j+\frac{1}{2}})} \ket{\{\sprimesi_i, \sprimet_{i-\frac{1}{2}} \}}
\end{split}
\end{eqnarray}
which is precisely the definition of $\KT^{\text{open}}$ on an open chain in \eqref{eq:KTopenchain}.

Having known that $\KT^{\text{open}}= \CN^\dagger U_{\text{DW}} \CN$, it is now clear how to find the Pauli operator representation of $\KT^{\text{open}}$. We first consider how the operators are mapped under the Kramers-Wannier transformation $\CN$,
\begin{eqnarray}
\begin{split}
    \widehat{\sigma}^x_{j-\frac{1}{2}} \CN  = 
    \begin{cases}
    \CN \sigma_{1}^z, &  j=1 \\
    \CN \sigma_{j-1}^z\sigma_{j}^z, & j=2, ..., L 
    \end{cases}, &\hspace{1cm}
    \widehat{\sigma}^z_{j-\frac{1}{2}} \CN = \CN \prod_{k=j}^{L} \sigma_k^x\\
    \widehat{\tau}^x_{j} \CN  = 
    \begin{cases}
    \CN \tau^z_{j-\frac{1}{2}} \tau^z_{j+\frac{1}{2}} , & j=1, ..., L-1\\
    \CN \tau^z_{L-\frac{1}{2}}, & j=L
    \end{cases}, &\hspace{1cm} \widehat{\tau}^z_{j} \CN = \CN \prod_{k=1}^j \tau_{k-\frac{1}{2}}^x.
\end{split}
\end{eqnarray}
Since $\CN$ is unitary according to Section \ref{sec:KWopen}, the above formula gives $\CN^\dagger \widehat{\sigma}^{x,z}_{j-\frac{1}{2}}\CN$ and $\CN^\dagger \widehat{\tau}^{x,z}_{j}\CN$. We then apply the above transformation to $U_{\text{DW}}$, and we get 
\begin{equation}\label{eq:KTopenchainfinal}
\begin{split}
    \CN^\dagger U_{\text{DW}} \CN &= \CN^\dagger (-1)^{\frac{1}{4}\sum_{j=1}^{L}(1-\widehat{\sigma}_{j-\frac{1}{2}}^z)(1-\widehat{\tau}^z_{j}) + \frac{1}{4}\sum_{j=1}^{L-1} (1-\widehat{\sigma}_{j+\frac{1}{2}}^z)(1-\widehat{\tau}^z_{j}) }\CN\\
    &= \CN^\dagger (-1)^{\frac{1}{4}\sum_{j=1}^{L-1}(1-\widehat{\sigma}_{j-\frac{1}{2}}^z \widehat{\sigma}_{j+\frac{1}{2}}^z)(1-\widehat{\tau}^z_{j}) + \frac{1}{4} (1-\widehat{\sigma}_{L-\frac{1}{2}}^z)(1-\widehat{\tau}^z_{L}) }\CN\\
    &= (-1)^{\frac{1}{4}\sum_{j=1}^{L-1}(1-\sigma_j^x)(1-\prod_{k=1}^{j}\tau^x_{k-\frac{1}{2}}) + \frac{1}{4} (1-{\sigma}_{L}^x)(1-\prod_{k=1}^{L}{\tau}^x_{k-\frac{1}{2}}) }\\
    &= (-1)^{\frac{1}{4}\sum_{j=1}^{L-1}(1-\sigma_j^x)(1-(\prod_{k>j}\tau^x_{k-\frac{1}{2}})(\prod_{k=1}^{L}\tau^x_{k-\frac{1}{2}}))  } (-1)^{\frac{1}{4}(1-\sigma_L^x)(1-\prod_{k=1}^{L}\tau^x_{k-\frac{1}{2}})  } \\
    &= (-1)^{\frac{1}{4}\sum_{j=1}^{L-1}(1-\sigma_j^x)(1-\prod_{k>j}\tau^x_{k-\frac{1}{2}})  } (-1)^{\frac{1}{4}\sum_{j=1}^{L}(1-\sigma_j^x)(1-\prod_{k=1}^L\tau^x_{k-\frac{1}{2}})  }\\
    &= \left(\prod_{j=1}^{L-1} \prod_{k> j} (-1)^{\frac{1}{4}(1-\sigma_j^x)(1-\tau^x_{k-\frac{1}{2}}) }\right) (-1)^{\frac{1}{4}(1-\prod_{j=1}^{L}\sigma_j^x)(1-\prod_{k=1}^L\tau^x_{k-\frac{1}{2}})  }.
\end{split}
\end{equation}
In the second equality, we used $(-1)^{\frac{1}{4}(1-\widehat{\sigma}^z_{j-\frac{1}{2}} + 1- \widehat{\sigma}^z_{j+\frac{1}{2}})(1-\widehat{\tau}_j^z)} = (-1)^{\frac{1}{4}(1-\widehat{\sigma}_{j-\frac{1}{2}}^z \widehat{\sigma}_{j+\frac{1}{2}}^z)(1-\widehat{\tau}^z_{j})}$. Indeed, when $\widehat{\tau}^z_j=1$, both sides are trivial. When $\widehat{\tau}^z_j=-1$, both sides are 1 if $(\widehat{\sigma}^z_{j-\frac{1}{2}}, \widehat{\sigma}^z_{j+\frac{1}{2}})=(1,1), (-1,-1)$, but are $-1$ if $(\widehat{\sigma}^z_{j-\frac{1}{2}}, \widehat{\sigma}^z_{j+\frac{1}{2}})=(1,-1), (-1,1)$.  The same trick is also applied to the fifth and last equality. The final expression is the $\KT$ in terms of Pauli operators, 
\begin{eqnarray}\label{eq:KT1}
\KT^{\text{open}}= \left(\prod_{j=1}^{L-1} \prod_{k> j} (-1)^{\frac{1}{4}(1-\sigma_j^x)(1-\tau^x_{k-\frac{1}{2}}) }\right) (-1)^{\frac{1}{4}(1-U_\sigma)(1-U_\tau)  }.
\end{eqnarray}
We note that this unitary operator is highly non-local.

\paragraph{Comparing $U_{\text{KT}}$ and $\KT^{\text{open}}$}

We finally proceed to relate $U_{\text{KT}}$ and $\KT^{\text{open}}$ on an interval, defined for spin-1 and spin-$\frac{1}{2}$ systems respectively. Using \eqref{eq:spin11/2},
the original Kennedy-Tasaki unitary operator $U_{\text{KT}}$ becomes
\begin{eqnarray}
U_{\text{KT}}= \prod_{i>j} e^{\frac{i \pi}{4} (\sigma^z_i+ \tau^z_{i-\frac{1}{2}})(\sigma^x_j+ \tau^x_{j-\frac{1}{2}})}.
\end{eqnarray}
Further applying \eqref{eq:basisrotation}, the $U_{\text{KT}}$ becomes
\begin{eqnarray}
U_{\text{KT}}= \prod_{i>j} e^{\frac{i \pi}{4} \sigma^z_i \tau^z_{j-\frac{1}{2}}(1- \tau^x_{i-\frac{1}{2}})(1-\sigma^x_j)} = \prod_{i>j} e^{\frac{i \pi}{4} (1- \tau^x_{i-\frac{1}{2}})(1-\sigma^x_j)}.
\end{eqnarray}
In the second equality above, we used the fact that the value of $\sigma^z_i \tau^z_{j-\frac{1}{2}} =\pm 1$, which only provides a sign, does not influence the exponent mod $2\pi$. This is precisely the first factor of $\KT$ in \eqref{eq:KT1}. Hence we have found that
\begin{eqnarray}\label{eq:KT2}
\KT^{\text{open}}= U_{\text{KT}} (-1)^{\frac{1}{4}(1-U_\sigma)(1-U_\tau)  }.
\end{eqnarray}
From \eqref{eq:KT2}, we find that the relation between $\KT^{\text{open}}$ and $U_{\text{KT}}$ depends on the symmetry sectors, labeled by the eigenvalues $(-1)^{u_{\sigma,\tau}}$ of $U_{\sigma, \tau}$. Concretely, 
\begin{eqnarray}\label{eq:KT3}
\KT^{\text{open}}= 
\begin{cases}
U_{\text{KT}}, & (u_\sigma, u_\tau)=(0,0), (0,1), (1,0)\\
-U_{\text{KT}}, & (u_\sigma, u_\tau)=(1,1).
\end{cases}
\end{eqnarray}
This completes the proof. 

To summarize, we managed to define the Kennedy-Tasaki transformation on a ring for both spin-1 and spin-$\frac{1}{2}$ systems, and have shown their equivalence. Both of them are non-unitary transformations, and satisfy non-invertible fusion rules, and showed that the latter implements $STS$ transformation. We also showed that when formulating them on an interval, the transformation $\KT^{\text{open}}$ for spin-$\frac{1}{2}$ systems and the original Kennedy-Tasaki transformation $U_{\text{KT}}$ for spin-1 systems are almost equivalent. Both of them are non-local and unitary transformations.

\section*{Acknowledgement}

We would like to thank Philip Boyle Smith, Weiguang Cao,  Yoshiki Fukusumi, Justin Kaidi, Ho Tat Lam, Emily Nardoni, Kantaro Ohmori, Shu-Heng Shao, Yuji Tachikawa, Masahito Yamazaki and Hong Yang for useful discussions. We also thank Yoshiki Fukusumi, Justin Kaidi, Zohar Nussinov, Yuji Tachikawa, and Hal Tasaki for comments on a draft.  L.L. is supported by Global Science Graduate Course (GSGC) program at the University of Tokyo. Y.Z. is partially supported by WPI Initiative, MEXT, Japan at IPMU, the University of Tokyo. This work was supported in part by MEXT/JSPS KAKENHI Grants No. JP17H06462 and No. JP19H01808, and by JST CREST Grant No. JPMJCR19T2.

\bibliographystyle{ytphys}
\baselineskip=.95\baselineskip
\bibliography{bib}

\end{document}